\documentclass[sigconf]{acmart}

\usepackage{subfig}
\usepackage{graphicx}
\usepackage{multirow}

\AtBeginDocument{%
  \providecommand\BibTeX{{%
    \normalfont B\kern-0.5em{\scshape i\kern-0.25em b}\kern-0.8em\TeX}}}

\setcopyright{acmcopyright}
\copyrightyear{2022}
\acmYear{2022}
\acmConference[ICPC '22]{30th International Conference on Program Comprehension}{May 16--17, 2022}{Virtual Event, USA}
\acmBooktitle{30th International Conference on Program Comprehension (ICPC '22), May 16--17, 2022, Virtual Event, USA}
\acmPrice{15.00}
\acmDOI{10.1145/3524610.3527906}
\acmISBN{978-1-4503-9298-3/22/05}

\acmPrice{15.00}
\acmISBN{978-1-4503-XXXX-X/18/06}

\begin{document}

\title{QuLog: Data-Driven Approach for Log Instruction Quality Assessment}

\author{Jasmin Bogatinovski}
\email{jasmin.bogatinovski@tu-berlin.de}
\affiliation{%
  \institution{Technical University Berlin}
  \streetaddress{Ernst Reuter Platz 7}
  \city{Berlin}
  \country{Germany}
}

\author{Sasho Nedelkoski}
\email{nedelkoski@tu-berlin.de}
\affiliation{%
  \institution{Technical University Berlin}
  \streetaddress{Ernst Reuter Platz 7}
  \city{Berlin}
  \country{Germany}
}

\author{Alexander Acker}
\email{alexander.acker@tu-berlin.de}
\affiliation{%
  \institution{Technical University Berlin}
  \streetaddress{Ernst Reuter Platz 7}
  \city{Berlin}
  \country{Germany}
}

\author{Jorge Cardoso}
\email{jorge.cardoso@huawei.com}
\affiliation{%
  \institution{Huawei Munich Reserch Center}
  \city{Munich}
  \country{Germany}
}

\author{Odej Kao}
\email{odej.kao@tu-berlin.de}
\affiliation{%
  \institution{Technical University Berlin}
  \streetaddress{Ernst Reuter Platz 7}
  \city{Berlin}
  \country{Germany}
}
\renewcommand{\shortauthors}{Bogatinovski, et al.}

\begin{abstract}
In the current IT world, developers write code while system operators run the code mostly as a black box. The connection between both worlds is typically established with log messages: the developer provides hints to the (unknown) operator, where the cause of an occurred issue is, and vice versa, the operator can report bugs during operation. To fulfil this purpose, developers write log instructions that are structured text commonly composed of a log level (e.g., "info", "error"), static text ("IP $\{\}$ cannot be reached”), and dynamic variables (e.g. IP $\{\}$). However, opposed to well-adopted coding practices, there are no widely adopted guidelines on how to write log instructions with good quality properties. For example, a developer may assign a high log level (e.g., "error") for a trivial event that can confuse the operator and increase maintenance costs. Or the static text can be insufficient to hint at a specific issue. In this paper, we address the problem of log quality assessment and provide the first step towards its automation. We start with an in-depth analysis of quality log instruction properties in nine software systems and identify two quality properties: 1) \textit{correct log level assignment} assessing the correctness of the log level, and 2) \textit{sufficient linguistic structure} assessing the minimal richness of the static text necessary for verbose event description. Based on these findings, we developed a data-driven approach that adapts deep learning methods for each of the two properties. An extensive evaluation on large-scale open-source systems shows that our approach correctly assesses log level assignments with an accuracy of 0.88, and the sufficient linguistic structure with an F$_1$ score of 0.99, outperforming the baselines. Our study highlights the potential of the data-driven methods in assessing log instructions quality and aid developers in comprehending and writing better code.
\end{abstract}

\ccsdesc[500]{Software and its engineering~Software testing and debugging}

\keywords{log quality,  deep learning, log analysis, program comprehension}

\maketitle

\section{Introduction}
Logging is important programming practice in modern software development, as software logs -- the end product of logging, are frequently adopted in diverse debugging and maintenance tasks. Logs record system events on arbitrary granularity and give insights into the inner-working state of the running system. The rich information they provide enables the developers and operators to analyze events and perform a wide range of tasks. Notable task examples relying on logs are comprehending system behaviour~\cite{Hassan2020}, troubleshooting~\cite{log2}, and tracking execution status~\cite{PinjiaHe2016}.

\begin{figure*}[!t]
\subfloat[An example of log instruction and generated message from the software system OpenStack.]{\includegraphics[width=0.33\textwidth]{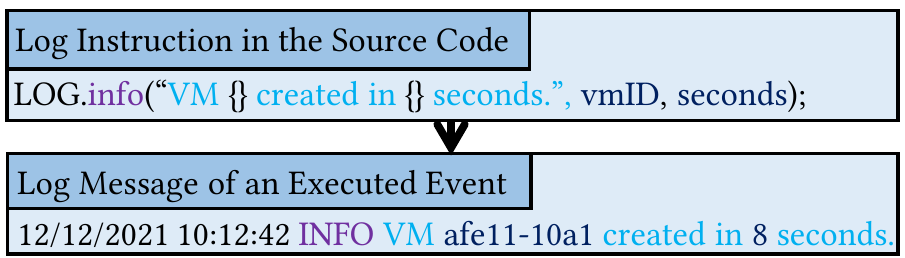}%
\label{fig1:exampl1}}
\hfill
\subfloat[Jira issue HDFS-4048: Example of wrong log level assignment and its fix.]{\includegraphics[width=0.33\textwidth]{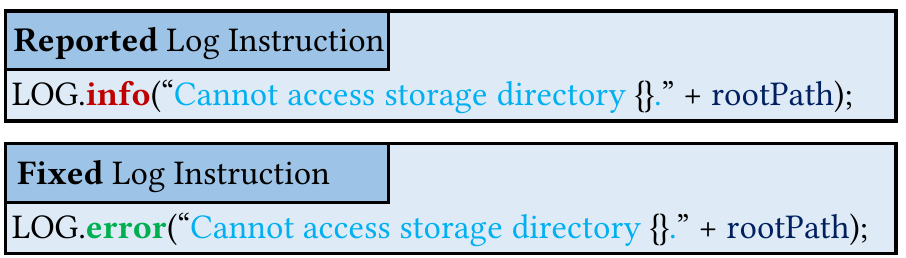}%
\label{fig1:exampl2}}
\hfill
\subfloat[Jira issue ZOOKEEPER-2126: Insufficient information/(linguistics) hurts event understanding. ]{\includegraphics[width=0.335\textwidth]{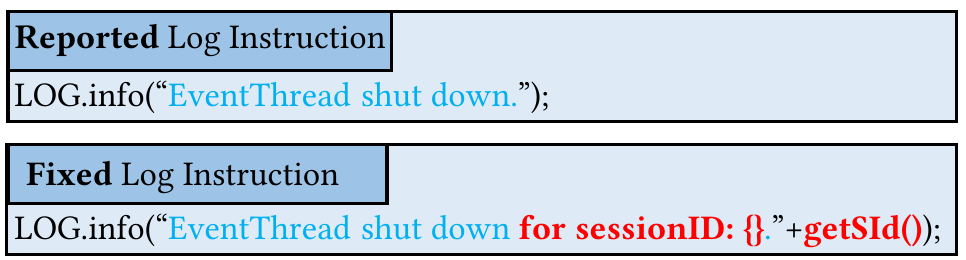}%
\label{fig1:exampl3}}
\caption{(a) Example of log instruction.  (b) \& (c) Examples of issues related to log instruction quality and their fixes.}
\label{fig}
\end{figure*}

Logs are textual event descriptors generated by log instructions in the source code. \figurename~\ref{fig1:exampl1} depicts an example of a log instruction and the log message (log for short) describing the executed system event. The log instructions are commonly composed of three parts 1) \textit{static text} describing the event (e.g., \texttt{VM \{\} created in \{\} seconds.}), 2) \textit{variable text} giving a dynamic information about the event (e.g., 8), and 3) \textit{log level} (e.g., info, error, warning), denoting the subjective developer opinion for the severity degree of the recorded event. The importance of log instructions makes them widely present within the source code. For example, HBase -- a popular Java software system, has more than 5k log instructions. Developers use diverse logging frameworks (e.g., Log4j~\cite{Log4J}) and logging wrappers (e.g., SLF4J~\cite{SLF4J}), which provide common logging features unifying the log instructions writing. 


Many companies are adopting logging frameworks and specify guidance to their developers on the \textit{quality} requirements when writing log instructions~\cite{whereToLogStudy2021}. The quality requirements define different properties of log instructions quality, such as 1) assignment of correct log levels, 2) writing static text with sufficient information (i.e., sufficient linguistic properties), 3) appropriate log instruction formats, and 4) correct log instruction placement within the source code~\cite{Chen2017}. The quality guidelines aim to align the expectations from the logs for both developers and operators that may work in different teams and use the logs for different activities. Since many maintenance tasks (e.g., tracing faulty activities, diagnosing failures, and performing root cause analysis) are frequently log-based~\cite{Hassan2020}, they directly depend on the quality of the logs, i.e., the quality properties of the log instructions. Therefore, evaluating the quality of the log instructions emerges as a relevant task. 

A central problem in this context is to write log instructions with sufficient quality. Recent studies on industrial~\cite{whereToLogStudy2021} and open-source software systems~\cite{Chen2017} suggest that developers make recurrent log-related commits during development. It means that writing quality log instructions for the first time, even with given quality guidelines, is not trivial. Additionally, the guidelines can be incomplete and do not cover every possible case. For example, in the Jira issue  LOG4J2-316~\footnote{\href{https://issues.apache.org/jira/browse/LOG4J2-3166}{https://issues.apache.org/jira/browse/LOG4J2-3166}}, a developer reported that the logging guidelines misguided him in proper usage of log levels. 

While the logging frameworks unify logging, they do not implement mechanisms to track the quality. Thereby, the decisions about the log instructions are purely human-centric, which can result in poor logging practices (e.g., wrong log level assignment or insufficient linguistic structure)~\cite{Hassan2018, manifesto}. For example, in the Jira issue HDFS-4048\footnote{\href{https://issues.apache.org/jira/browse/HDFS-4048}{https://issues.apache.org/jira/browse/HDFS-4048}} (depicted in \figurename~\ref{fig1:exampl2}), the wrong log level of the instruction \textit{LOG.info("Cannot access storage directory " + rootPath);} resulted in a long time for localization of the failure. The developer used the log levels "error" and "warning" for log-based failure localization, but the event initially was logged on log level "info", not "error". Similarly, in the Jira issue ZOOKEEPER-2126\footnote{\href{https://issues.apache.org/jira/browse/ZOOKEEPER-2126}{https://issues.apache.org/jira/browse/ZOOKEEPER-2126}} (depicted in \figurename~\ref{fig1:exampl3}), the log instruction has insufficient information about which EventThread is terminated. As reported by the developers, it becomes confusing when a new EventThread is created before terminating the previous one. The lack of a session identifier was pointed to as the main concern. The problem is resolved by adding additional words in the static text to give minimal information about the event which can be understood/comprehended by the developers. Notably, in linguistic terms, this means enriching the linguistic structure of the static text. The aforenamed issues are not isolated events. Previous works on logging practices~\cite{Hassan2018, Chen2017} suggest that it is surprisingly common for the log levels to be over/under-estimated or the logs to have missing or excessive information. These problems are particularly challenging in complex software, with many different components developed by multiple developers located in diverse geographical regions (e.g., systems like OpenStack). It requires non-trivial knowledge and experience to construct an understandable description of the event, estimate the log levels of the instruction or conduct quality logging practices in general. Although the human-centric approach in log quality assessment is the golden standard, the aforenamed challenges imply the need for an automatic approach.

In this paper, we address the log quality assessment problem. Our goal is to develop an approach to automatically assess the quality of log instructions from an arbitrary software system. Such an approach is challenged by the heterogeneity of the software systems, the unique writing styles of developers, and different programming languages. They limit the set of testable quality properties. For example, the different syntax of the nearby code from two programming languages (e.g., Java and Python) questions the applicability of log instruction placement methods on arbitrary system~\cite{whereToLogStudy2021}. To find the empirically testable properties, we performed a preliminary manual study on the properties of the log instructions from nine open-source systems. We identified two such quality properties -- 1) \textit{log level assignment} assessing the correctness of the log level, and 2) \textit{sufficient linguistic structure} assessing the minimal linguistic richness of the static text necessary for verbose event description. Through the preliminary study, we find that the static text of the log instruction is sufficient in assessing the two properties. Therefore, the log quality assessment is done on the static text of the log instructions, independent of the other properties of the source code (e.g., nearby code structure). This makes the log quality assessment system-agnostic. The observed dependencies between the static text on one side, and the log levels and linguistically sufficient labels on the other side allow the application of data-driven methods. Ultimately enabling the automation of the log instruction quality assessment.

Based on our observations, we propose QuLog as an approach to automatically assess log instructions quality. QuLog trains two deep learning models from the log instructions of many software systems and appropriate learning objectives to learn quality properties for the log levels and sufficient linguistic structure of static texts. To capture diverse developers logging styles, we trained the models on a carefully constructed log instruction collection with expected good quality practices similar as in related work~\cite{Chen2017}. By adopting an approach from explainable AI, we further implemented a prediction explainer to show why the models make certain predictions, which serve as additional feedback for developers. Our experimental results show that QuLog achieves high performance in assessing the two quality properties outperforming the baselines. The prediction explainer has a low error for the correct prediction reason suggestion. Thereby, the experiments show that QuLog helps to assess the log instructions quality while giving useful suggestions for their improvement. 

In a nutshell, our contributions summarize as:
\begin{enumerate}
    \item We performed a manual analysis on the quality log properties of the log instructions on nine software systems and identified 1) log level and 2) sufficient linguistic structure assessments as two empirically testable properties.
    \item We implemented a novel approach for automatic system-agnostic log quality assessment named QuLog, which uses deep learning and explainable AI methods to evaluate the two properties.
    \item We experimentally demonstrate the usefulness of our approach in automatic system-agnostic log quality assessment, which achieves high accuracy for log level assignment (0.88) and a high F$_1$ on sufficient linguistic structure (0.99) assessments. 
    \item We open-source the code, datasets and additional experimental results in the code repository~\cite{githubrepoVisible}.
\end{enumerate}

\section{Log Instruction Quality Assessment}~\label{sec2}
\subsection{Log Instruction Quality Properties}~\label{sec21} To assess the quality of the log instructions, we examined literature studies on logging practices. We identified two views: explicit (or developers), and implicit (or operators). The explicit view is related to (a.1) correct log level assignment, (a.2) comprehensive content of the static text and parameters, and (a.3) correct log instruction placement~\cite{WeviShen2018}. The implicit view is related to the operators' expectations for the quality of the logs. By observing the properties of the logs, we can \textit{implicitly} reason about the quality properties of the log instruction. For the implicit view, there are four properties~\cite{manifesto}, given as follows: (b.1) \textit{trustworthiness} - refers to the valid meta-information of the log (e.g., correct log level), (b.2) the \textit{semantics/linguistic} of the log - relates the word choice in verbose expression of the event, (b.3) \textit{completeness} - reflects the co-occurring of logs to describe an event, and (b.4) \textit{safeness} - refers to the log content being compliant with user safety requirements.

Since our goal is to provide an automatic log instruction quality assessment, we first examine the feasibility of automatically evaluating the properties. We observed that some of the properties (i.e., correct log level assignment and linguistic evaluation) depend and can be assessed just from the content of log instructions. Therefore, they can be evaluated irrelevant to the structure of the source code and the remaining logging practices. To verify our observation, we made a preliminary study of nine open-source systems, with presumably good logging practices (similarly as in related works~\cite{DeepLV, Chen2017, PinjiaHe2016}). \tablename~\ref{tab:deeplv} enlists the properties of the used systems. We select these systems because they serve many industries, are being developed by many experienced developers, and consequently, the logs have fulfilled their purpose in debugging and maintenance.

\begin{table}[!h]
\caption{Overview of the studied systems}
\label{tab:deeplv}
\begin{tabular}{l|c|c|c}
\hline
\multicolumn{1}{l|}{Software System} & Version & \multicolumn{1}{c|}{LOC} & NOL  \\ \hline
Cassandra                             & 3.11.4  & 432K                     & 1.3K \\ 
Elasticsearch                         & 7.4.0   & 1.50M                    & 2.5K \\ 
Flink                                 & 1.8.2   & 177K                     & 2.5K \\ 
HBase                                 & 2.2.1   & 1.26M                    & 5.5K \\ 
JMeter                                & 5.3.0   & 143K                     & 1.9K \\ 
Kafka                                 & 2.3.0   & 267K                     & 1.5K \\ 
Karaf                                 & 4.2.9   & 133K                     & 0.7K \\ 
Wicket                                & 8.6.1   & 216K                     & 0.4K \\ 
Zookeeper                             & 3.5.6   & 97K                      & 1.2K \\ \hline
\end{tabular}

{\raggedright \small{Note: LOC and NOL stand for the number of code and log lines accordingly.\par}}
\end{table}

\subsection{Empirical Study}

\subsubsection{Log Level Assignment}~\label{esllq} We assume that the static text of the log instruction has relevant features for log level assessment. Intuitively, when describing an event with the "error" log level, the static text commonly contains words like "error", "failure", "exit", and similar. Whenever these words occur within the static text, it is more likely that the level is "error" than "info". To verify our assumption, we considered an approach from information theory that defines the amount of uncertainty of information in a message~\cite{informationtheory}. In our case, we analyze the relation of word groups (n-grams, $n=\{1, 2, 3, 4, 5\}$) from the static text in relation to the log level. For all the n-gram groups, we try to identify the log level using n-grams from the given static text of the log instruction. At first, given an n-gram, there is high uncertainty for the assigned log level. As we receive more information about the n-gram, we update our belief for its commonly assigned log level, reducing the entropy (uncertainty) associated with the n-gram. To measure the uncertainty, we used Normalized Shanon's entropy~\cite{ShanonEntropy}. We calculated the log level entropy for each n-gram from all the log instructions of the nine software systems and reported the key statistics for the distribution.

\begin{table}[!h]
\caption{Log level assignment empirical study results.}
\label{loglevelentropy}
\begin{tabular}{c|c|c|c|c|c}
\hline
                                                                      & Min  & 1st Qu. & Median & 3rd Qu. & Max  \\ \hline
\begin{tabular}[c]{@{}c@{}}Average Entropy\end{tabular} & 0.00 & 0.00    & 0.00   & 0.56    & 0.91 \\ \hline
\end{tabular}
\end{table}
\tablename~\ref{loglevelentropy} summarizes the n-grams entropy distribution. It is seen that the majority of the static text of the log instructions have low entropy. Specifically, more than 50\% (the median) of the static texts have zero entropy -- the n-grams
appear on a unique level. Therefore, \textit{the static text has relevant features useful to discriminate the log levels}, verifying our assumption.


\subsubsection{Linguistic Quality Assessment}~\label{lqa}
A quality log instruction should describe the event concisely and verbosely~\cite{Chen2017}. From a general language perspective, complete and concise short texts (following the maxims of text quantity and quality) have a minimal linguistic structure (e.g., usage of nouns, verbs, prepositions, adjectives)~\cite{Finegan2014}. Under the term log linguistic structure, we understand the representation of the static text by general linguistic properties such as linguistic concepts (e.g., verbs, nouns, adjectives etc.). For example, in the Jira issue ZOOKEEPER-2126 (depicted in \figurename~\ref{fig1:exampl3}), the static text "EventThread shut down." linguistically is composed of "noun verb particle". Owning to the shared properties of the general English language and language used in log instructions~\cite{PinjiaHe2018}, we assume that an informative event description also has a minimal linguistic structure. The following example explains our intuition for the assumption. In the aforenamed Jira issue, developers reported that the event information is insufficient. This issue is resolved by static text augmentation with additional linguistic properties, i.e., "EventThread shut down for session: \{\}", linguistically composed of "noun verb particle preposition noun: -LRB- -RRB-" (where "-LRB- -RRB-" denote brackets). Linguistically speaking, the static text with insufficient linguistic structure is transformed into static text with sufficient structure, improving the event comprehension. Examples of other Jira issues related to sufficient linguistic quality can be found in Appendix~\ref{additionalEvidence}. 

To validate our assumption, we performed the following experiment. For the static text of each log instruction, we first extract their linguistic structure. To do so, we use part-of-speech (POS) tagging -- a learning task from NLP research. It allows extraction of the linguistic structure of the static text by linking the words to an ontology of the English language (OntoNote5). We choose spacy implementation of POS tagging because its models have high performance on the POS tagging task (>97\% accuracy score)~\cite{spacy}. Second, we group the extracted linguistic structures such that the static text with the same linguistic group are placed together. Afterwards, the linguistic groups of the raw static text are evaluated by two experienced developers answering the research question: "Does the static text from the examined linguistic group contains minimal information required to comprehend the described event?". This question evaluates our assumption that the quality and self-sustained static text has a minimal linguistic structure aligned with expert intuition for a comprehensible event description. 

\begin{table}[!h]
\caption{Linguistic quality assessment preliminary study}
\label{tabling}
\begin{tabular}{c|c|l}
\hline
\multicolumn{1}{l|}{Linguistic Group} & \begin{tabular}[c]{@{}c@{}}Total Log\\ Instructions\end{tabular} & \multicolumn{1}{c}{\begin{tabular}[c]{@{}c@{}}Static Text\\ (Example)\end{tabular}} \\ \hline
VERB NOUN                              & 106                                                              & serialized regioninfo                                                                \\ \hline
VERB                                   & 67                                                               & deleted                                                                              \\ \hline
VERB PUNCT                             & 49                                                               & interupted *                                                                         \\ \hline
NOUN                                   & 47                                                               & return                                                                               \\ \hline
NOUN NOUN                              & 41                                                               & updating header                                                                      \\ \hline
\end{tabular}
\end{table}

\tablename~\ref{tabling} gives the top-5 frequent linguistic groups alongside representative examples. In total, we found 5.9K linguistic groups from the studied systems. Then, we randomly sample 361 groups based on a 95\% confidence interval and a 5\% confidence level~\cite{StatNut}. The sampling is stratified over the nine systems. The two human experts identified 24 linguistic groups with insufficient linguistic structure. The agreement between the two experts assessed by Cohen's Kappa score is sustainable (0.72)~\cite{highKappa}. The high score values show mutual agreement between the experienced developers concerning the relation between comprehensible event information within the static text and its linguistic structure. Therefore, \textit{the linguistic structure of the static text} is useful in representing the minimal informative description of the log instruction. 

\subsubsection{Other Quality Properties} The remaining quality properties (i.e., relevant variable selection, log instruction placement, safeness, and completeness) depend on the different programming languages, design patterns, and other source code structures. These properties are challenging for assessment because of the heterogeneity of software systems and the ways programming languages organize the source code. For example, the safeness property requires reasoning across a complex chain of method invocations (e.g., in the issue \href{https://nvd.nist.gov/vuln/detail/CVE-2021-44228}{CVE-2021-44228}  the bug allows execution of any Java method through the log instruction from an LDAP server hurting the logging safeness). Identifying safeness in this context requires a deep understanding of potential method invocation chains, which does not even require the method's presence within the source code, i.e., requiring human involvement. The latter is against our effort in automatic log quality assessment. Due to the identified relationships between the static text and log level and sufficient linguistic structure on one side, and the dependence of the other quality properties on the remaining parts of the source code on the other side, we consider the log quality assessment in the narrower sense, composed of the former two quality properties.

\section{QuLog: Automatic Approach for Log Instruction Quality Assessment}~\label{sec3}
Inspired by our findings in the preliminary study, we propose an approach for automatic system-agnostic log instruction quality assessment. We formulate the problem in the scope of 1) evaluating the correct log level assignment and 2) evaluating the sufficient linguistic structure of the log instructions. Given the static text of the log instruction, we apply deep learning methods to learn static text properties concerning the correct log level and sufficient linguistic structure. By training the models on systems with quality logging properties, they learn information for the log level and sufficient linguistic structure qualities. Comparing the predicted log levels and the log levels assigned by developers allows a statement on the log level quality: the less deviation, the better the quality. Similarly, the sufficient linguistic structure incorporates properties of comprehensible log instructions, and its predictions directly are used to assess linguistic quality.

\begin{figure*}[!h]
\centering
\includegraphics[width=0.7\textwidth]{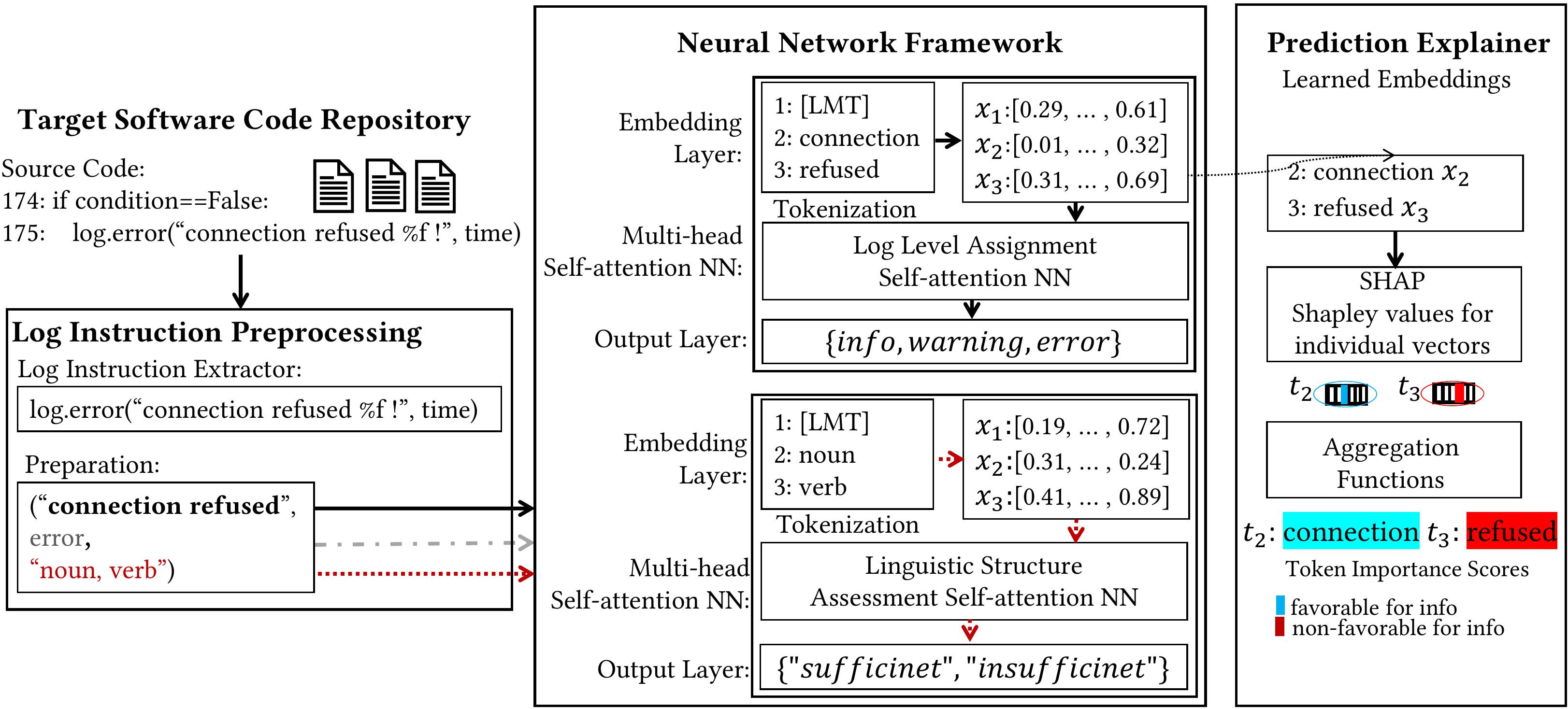}
\caption{Internal architectural design of QuLog}
\label{fig:encoderBlock}
\end{figure*}

Figure~\ref{fig:encoderBlock} illustrates the overview of the approach, named QuLog. Logically, it is composed of (1) \textit{log instruction preprocessing}, (2) \textit{deep learning framework} and (3) \textit{prediction explainer}. The role of the \textit{log instruction preprocessing} is to extract the log instructions from the input source files and process them into a suitable learning format for the deep learning framework. The \textit{deep learning framework} is composed of two neural networks (one for each of the two quality properties). The neural networks are trained separately on the two tasks. After training, the networks learn discriminative features for the log instructions with different log levels and a sufficient linguistic structure. The \textit{prediction explainer} explains a certain prediction. Specifically, given the static text of the log instruction and predicted log level, it shows how different words contribute to the model prediction. 

QuLog has two operational phases: offline and online. During the offline phase, the parameters of the neural networks and explanation part are learned on representative data from other software systems. This training procedure allows learning diverse developers writing styles, important for generalization. The learned models are stored. In the online phase, the source files of the target software system are given as QuLog's input. QuLog extracts the log instructions, the static texts and log levels, proceeding them towards the loaded models. As output QuLog provides the predictions for the log levels, sufficient linguistic structure, and prediction explanations as word importance scores. Therefore, QuLog serves as a standalone recommendation approach to aid developers in improving the quality of the log instructions. The developers may reconsider improving the log instructions given QuLog's suggestions or reject them. In the following, we delineate the details of the three QuLog's components.

\subsection{Log Instruction Preprocessing}
The purpose of the log instruction preprocessing is twofold. First, it extracts the log instructions from the source files. Second, it parses the log instructions to separate the static text and the log level from the remaining instruction parts. In addition, the static text is processed by the linguistic features extractor, to obtain its linguistic structure representation. These operations are performed by two modules, namely (1) log instruction extractor and (2) log instruction preparation, described in the following. 

\sloppy{\subsubsection{Log Instruction Extractor} The extractor module extracts the log instructions from the source code of the software system. To that end, it iterates over all of the source files in the target software's source code and applies regular expressions to find all log instructions. Considering the diversity of the programming languages, developers writing styles, and the lack of adoption of logging practices challenges the extraction process. The output of the extraction module is a set of log instructions of the input software system. Although our goal is to help developers in writing correct log levels, we restrain ourselves on three levels ("info", "warning", and "error"). The log levels "trace" and "debug" provide detailed information for the inner workings, most commonly used by developers. By studying the n-grams frequency for individual log levels, we found that there is a large overlap between the used vocabulary in "info" and "trace"/"debug" levels. This can significantly impair the performance of the data-driven methods when automatically assessing the quality of all log levels simultaneously. In addition, related work reports this scenario practically useful when different stakeholders examine logs. For example, operators care more for the high severity levels (i.e., "info", "warning", "error")~\cite{DeepLV}.}

\subsubsection{Preparation} The goal of the preparation module is to prepare the data in a suitable learning format. As input, it receives the set of log instructions from the extractor. The preparation module first iterates over the log instructions and separates the static text of the log instructions from the log level. 
The diverse programming languages use different names for the log levels. For example, Log4j (a Java's logging library) uses the tag $"warn"$ for warning logs, while the default Python logging library uses the tag $"warning"$. To that end, the preparation submodule unifies the levels for all log instructions. To the static text of the log instruction, we apply Spacy~\cite{spacy} for preprocessing. We split the words using space and camel cases. We preprocess the static text by following text preprocessing techniques, including: remove all ASCII special characters, removing stopwords from the Spacy English dictionary and applying lower case transformation of the words~\cite{28}. Once processed, we give the static text as input to a pre-trained POS tagging model from Spacy. We extract the $pos$ tag of each word from the static text to create its linguistic structure. Finally, the output of the preparation module is a set of tuples, where each tuple is composed of the static text of the log instruction, the linguistic structure of the static text and the log level.

\subsection{Deep Learning Framework}\label{sec:section32}

\subsubsection{Overall Architecture} QuLog has two independent neural networks to assess the two quality properties. They share the same architectural design and are composed of embedding layer, encoder network from Transformer architecture~\cite{Transformer} and output layer. To make the description easy to follow, we explain the working principle for the log level assignment. Alongside it, in parenthesis, we give the mapping for sufficient linguistic structure assessment. Given the preprocessed static text (linguistic structure) at the input, the embedding layer learns numerical vector representation of the individual words (linguistic categories), we referred to as tokens, following a distributed learning representation paradigm~\cite{distributedRepresentations}. The vector embeddings of the tokens are numerical features in a suitable learning format for the network. We then use the encoder of the Transformer architecture to learn relationships between the vector embeddings of the input tokens from the embedding layer and the log levels (sufficient/insufficient linguistic structure). The output from the encoder layer is a vector embedding of the static text (linguistic structure). After that, the output layer predicts log level (sufficient/insufficient linguistic structure) from the encoder layer's output.

\subsubsection{Embedding Layer} The embedding layer receives the preprocessed instructions as input. We first transform the static text (linguistic structure) from a sequence of words to a sequence of tokens/indices. \figurename~\ref{fig:encoderBlock} gives an example of this transformation. It enables the transition from textual into a numerical format as a prerequisite to applying neural networks. We further prepend the tokenized static text/(linguistic structure) with a special token we refer to as Log Message Token ([LMT]). Note that this is an important detail we discuss further when describing the neural networks. Since the static texts can be of different lengths, while the neural network requires fixed-size input, we specify a hyperparameter $max\_len$ to unify the lengths. The shorter static texts (linguistic structures) are appended with a special pad token ([PD]), while the longer ones are truncated at $max\_len$ value. The embedding layer maps the input tokens into a numerical vector representation, such that each unique token is assigned a specific vector. In QuLog, these embeddings are learned during training, and the vector embeddings are adjusted to preserve contextual properties (e.g., frequently co-occurring words for a certain log level). These vectors can also be obtained from general-purpose language models (e.g., BERT~\cite{devlin2018bert}). 

\subsubsection{Transformer Neural Network}
We model the dependencies between the tokens and the two quality properties with nonlinear parametric functions represented as neural networks. As a suitable architecture, we identified the encoder of the Transformer~\cite{Transformer} architecture. It provides state-of-the-art results in many NLP tasks (e.g., sentiment analysis, translation)~\cite{GPT3}. By pointing to the similarities between the static text of logs and natural language~\cite{PinjiaHe2018}, we further justify our design of choice. The encoder implements a multi-head self-attention mechanism that exploits higher-order relations between tokens within the static text (beyond n-gram counting). This property captures discriminative features between the words (linguistic concepts) and the different contexts they appear in, relating them to the appropriate log levels (sufficient linguistic structures). During training, the token embedding vectors and the network parameters are updated via backpropagation~\cite{Linnainmaa1976}. At the output of the encoder, we provide the vector embedding of the [LMT] token. Due to the architectural design, the vector of the [LMT] token attends over all the other token vector embeddings during training. This allows summarizing the most relevant information from the input concerning the log level (sufficient linguistic structure). Therefore, it embeds diverse contexts preserving the properties of the static text (linguistic structure). The number of heads in the multi-head self-attention, the number of layers and model size are three hyperparameters of the network.

\subsubsection{Output Layer}
The output layer is a three-dimensional linear layer for predicting the log level (two-dimensional for sufficient linguistic structure). It accepts the [LMT] vector embedding and applies a linear transformation. Each of the output dimensions corresponds to one of the log levels (i.e., "info", "warning", "error") or one of the two linguistic structures qualities (i.e., sufficient and insufficient). We apply a softmax function at the output neurons to produce score estimates. Each neuron gives a score estimate for the corresponding class (i.e., a number between 0-1 indicating class relevance for the given input). The scores give insights into the model's confidence for the log level (sufficient linguistic structure) prediction. As a class prediction, we considered the class related to the neuron with the highest score.

\subsection{Prediction Explainer}\label{sec:section33}
The prediction explainer aims to explain why a trained model makes certain predictions. It augments QuLog's output by pointing out the specific tokens most likely contributing to the prediction. They can serve as suggestions to developers for where the static text potentially can be altered. The relevant details of the explanation module are given in the following.

\subsubsection{SHAP} Prediction explainer leverages SHAP~\cite{shap} (Shapley additive explanations) -- an approach from explainable artificial intelligence. In general, SHAP calculates feature importance scores (how relevant is a feature for the prediction) by defining the problem as a coalitional game between the features. The goal is to find the so-called Shapley values for each feature defined as the fairest distribution of the "payout" (as importance score) for the prediction. The larger the value, the more important is the feature towards the prediction. The signs of the Shapley values show the feature's prediction favorableness (or non-favorableness for negative sign).

\subsubsection{Implementation} We used the original SHAP implementation with the default values for its parameters~\cite{shapgithub}. One required parameter of SHAP is a differentiable learning model (a model with gradients calculated for each network layer). To apply SHAP, we used the trained encoder network as input. While the explanation procedure is applicable for both quality properties, we described just the log level prediction explainer because of the intuitive meaning of the importance scores concerning the predictions. 
\begin{figure}[!t]
\centering
\includegraphics[width=\columnwidth]{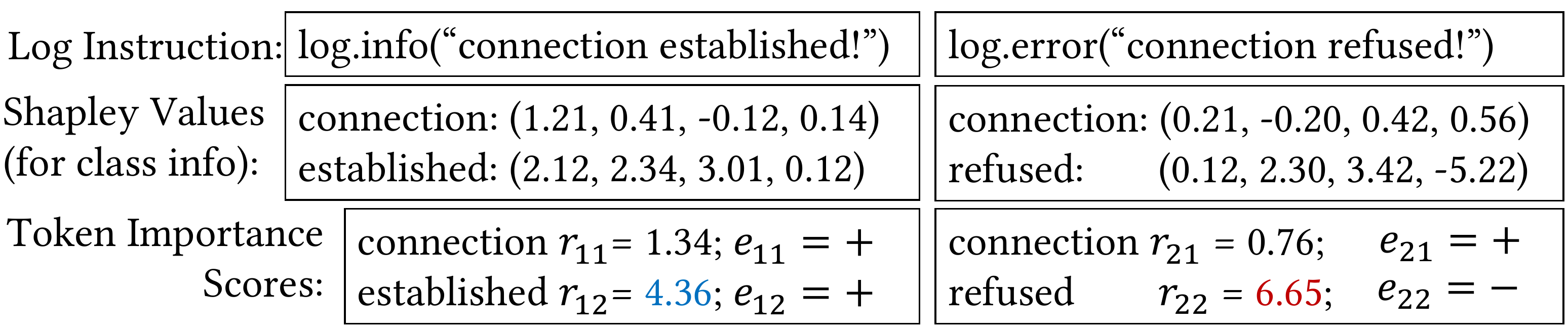}
\caption{Prediction Explainer working procedure example}
\label{fig:shapexp}
\end{figure}
\subsubsection{Token Importance Scores.} The Shapely values are calculated for each neuron of the input vector embeddings.  \figurename~\ref{fig:shapexp} illustrates a running example. There are two log instructions $l_1:$ "Connection established!" with the level "info", and  $l_2:$ "Connection refused!" with the level "error". After running the two instructions through SHAP, each token is assigned with a vector of Shapley values (with size $d=4$). However, to reason about the influence of the token in unity, we express the token importance as a single number. We refer to this as a \textit{token importance score}. To calculate the token importance score, we aggregate the individual Shapely values for each token. The \textit{token importance score} has two parts: 1) intensity and 2) sign. The intensity shows the influence strength of the token for the prediction. The sign shows the token direction influence for the prediction (favouring or not-favouring a decision). After experimentation with different aggregation functions, we find that the second norm of the Shapely value vector and the sign of the Shapley value with the highest absolute score, are suitable for intensity and sign aggregation functions. Formally, they are given in Eq.~\ref{ShapelyMagnitute} and Eq.~\ref{ShapelySign}.
\begin{equation}
    r(t_{ji})=||S_{ji}(t_{ji})||_2^2
    \label{ShapelyMagnitute}
\end{equation}
\begin{equation}
    e(t_{ji})=sign[\max_{k}|S_{jik}(t_{ji})|]
    \label{ShapelySign}
\end{equation}
where $r(t_{ji})$ is the token ($t_i$) importance score intensity, $e(t_{ji})$ is the token importance sign for the log instruction $l_j$. $S_{jik}$ denotes the Shapely value for the "$k$-th" position of the "$i$-th" token of the log instruction static text (i.e., in the example the token "refused" has a Shapely value $S_{224}=-5.22$).



In the example, the difference between the two instructions distinguishing the levels "info" from "error" is in the second token. We first calculate the individual Shapely values and then calculate the intensity and sign of the token importance scores. As seen by the score values, the following inequalities hold $r_{12}>r_{11}, r_{22}>r_{21}$. The second token in both of the instructions ("established", "refused") has greater intensity compared to the first ("connection"), thereby, contributing more to the model prediction. Additionally, the token signs show that the token "established" is favourable for the class "info" ($e_{12}=+$),  while the token "refused" is not-favourable for the class "info" ($e_{22}=-$). Therefore, if there is a discrepancy between the developers' decision on the log level and QuLog's log level assignment, the developer examines the highlighted word, e.g., "refused", and considers changing either the level or the word. That way, QuLog automatically aids developers in improving the log quality. The output of the explanation module is an ordered list of tokens, ordered by their intensity (from highest to lowest), examined by developers in that order.

\section{Experimental Evaluation}~\label{sec4}
\subsection{Experimental Setup}
\subsubsection{Code Repository Collection} Alongside the studied software systems, we collected log instructions from 100 other systems from GitHub. To collect this data, we crawled GitHub and searched for different systems from the following topics: Java, Python, Angular, Ruby, and PHP, selecting 7039 systems. Additionally, we collected the number of GitHub stars for each system. Similar to previous works~\cite{PinjiaHe2018}, we assumed that the number of stars is a good indicator for the quality of logging and considered the top 100 in the experiments. The usage of this data is described in the corresponding evaluation setting where used. 

\subsubsection{Evaluation Criteria} To evaluate QuLog, we used several evaluation criteria. First, we describe the criteria evaluating exact predictions. For the multi-class evaluation, we used \textit{accuracy}. It gives the percentage of correct predictions out of all of the predictions. Due to the imbalances of the target classes (e.g., different systems have a diverse number of "error", "warning", and "info" instructions), accuracy can be misleading~\cite{Gibaja2014}. Therefore, we further considered \textit{precision}, \textit{recall} and \textit{F$_1$} scores. \textit{Precision} evaluates the fraction of correct predictions out of all class predictions. \textit{Recall} evaluates the correct predictions out of all true class predictions. \textit{F$_1$} is the harmonic mean of the precision and recall evaluating the trade-off between correct class predictions and the miss-classifications~\cite{Gibaja2014}. Additionally, we considered \textit{specificity} in the binary classification setting. It is the measure of correct predictions of the negative class. The aforenamed criteria take values within the 0-1 range, and a higher value indicates better performance. The evaluation of the prediction explainer is done with the $error@k$ score. It measures the number of incorrect predictions when k-degrees of freedom are considered~\cite{Thorsten2005}. The smaller values indicated better performance. We additionally considered the \textit{AUC}~\cite{auc} score. \textit{AUC} is the area under the \textit{ROC} (receiver operating characteristic) curve that plots the true positive against the false-positive rates. It is bounded in the 0-1 range, with a high value indicating better performance. The AUC value of 0.5 indicates a model performing not better than a random guess.

\subsection{Log Level Assignment Evaluation} 
We first evaluate the performance of QuLog on the log level quality assessment. We split this evaluation into two parts. First, we compare QuLog against baselines. Second, we evaluate QuLog's performance on a few instances of the problem of log level assignment. The motivation for this evaluation type on one side is to examine the performance of QuLog against baselines, and on the other side, to identify problem instances where the inherited imperfect performance of data-driven approaches will not overwhelm developers with many incorrect predictions. The latter is relevant for QuLog's practical usability.

\subsubsection{Comparison against baselines} In the evaluation of QuLog against baselines, we considered two QuLog models. They have the same architectural design but differ in the input data used to train them. The first model, we referred to as QuLog-8, is trained on data from eight software systems listed in \tablename~\ref{tab:deeplv}. Since these systems are characterized by good logging practices, we assume that the majority of the log levels are correctly assigned, similar as in previous studies~\cite{DeepLV}. This accounts for the quality of the learning data. The second model for log level assignment we call QuLog* is trained on the collection of 100 GitHub systems. While QuLog* does not account very rigorously for the instruction quality (despite the pseudo indicator of having many stars), it enables testing for cross-software usefulness of the static text in log level assignment. As such, it aligns with the system-agnostic nature of QuLog. This is important in scenarios of log quality assessment where the software system is in the initial development stage, and there are not many log instructions for training a model. As an evaluation dataset, we considered the log instructions from one of the nine systems listed in \tablename~\ref{tab:deeplv}, such that the instructions from the evaluation dataset are never seen during training the model, preventing data leakage.  


\textit{Experiment Design.} We compare QuLog against two baselines: DeepLV~\cite{DeepLV} and Support Vector Machines (SVM)~\cite{Vapnik1995}. DeepLV addresses the problem of log level assignment as an ordinal regression and trains LSTM -- a deep-learning architecture, on features extracted from the log instruction. It is reported as the current best performing method for log level assignment. SVM is a popular multi-class classification method trained on the vector representation of the static text from general-purpose language models  (BERT)~\cite{devlin2018bert} previously used for log level assignment~\cite{Ann2019}. The hyper-parameters of the baseline methods are set to the recommended values by the authors. As evaluation criteria, we used AUC and accuracy following literature standards~\cite{DeepLV, Hassan2018}. Regarding the considered hyper-parameters for QuLog's log level architecture, we set the number of heads to two, the model size to 16, the number of layers is set to two, and the maximal number of tokens to $max\_len=50$. For training the model we used Adam optimizer~\cite{adam} with learning rate$10^{-4}$ and hyperparameters $\beta_1=0.9, \beta_2=0.99$. The \textit{batch size} was set to 2048.

\textit{Results and discussion.}
\tablename~\ref{tab:loglvlbaselines} gives the results of the evaluation of QuLog against baselines. We first compare the QuLog-8 model against the two baselines. Following the AUC criteria, it is seen that QuLog-8 achieves the best performance for all of the nine systems. Since AUC evaluates how good a model is in predicting the true level (e.g., "error") as correct (as "error") rather than predict incorrect level as the true one (e.g., "warning" instead of "error"),  it means that QuLog-8 can discriminate the different log levels better. However, AUC evaluates scores instead of exact decisions for a particular log level. By deciding for a log level (i.e., maximal score estimate as a class prediction), we assign an actual log level for the given input instruction. To evaluate the correctness of the log level decisions, we use accuracy. Comparing QuLog-8 against DeepLV it is seen that it is outperforming it in 8/9 systems while failing to do so in 1/9 (HBase) systems. Comparing QuLog-8 against SVM shows that QuLog performs better on 6/9 datasets, performs worse in 2/9 systems and ties on 1/9 (Flink). The evaluation criteria show that our approach is useful in assessing the correctness of log levels for the considered systems outperforming the baselines.

\begin{table}[!t]
\caption{Evaluation on log level quality assessment}
\label{tab:loglvlbaselines}
\resizebox{\columnwidth}{!}{%
\begin{tabular}{l|cccc|cccc}
\hline
              & \multicolumn{4}{c|}{AUC}                                                                      & \multicolumn{4}{c}{Accuracy}                                                                 \\ \hline
Systems       & \multicolumn{1}{c|}{QuLog-8} & \multicolumn{1}{c|}{DeepLV} & \multicolumn{1}{c|}{SVM}  & QuLog* & \multicolumn{1}{c|}{QuLog-8} & \multicolumn{1}{c|}{DeepLV} & \multicolumn{1}{c|}{SVM}  & QuLog* \\ \hline
Cassandra     & \multicolumn{1}{c|}{0.94}  & \multicolumn{1}{c|}{0.78}   & \multicolumn{1}{c|}{0.80} & 0.96   & \multicolumn{1}{c|}{0.63}  & \multicolumn{1}{c|}{0.61}   & \multicolumn{1}{c|}{0.64} & 0.67   \\ 
Elasticsearch & \multicolumn{1}{c|}{0.93}  & \multicolumn{1}{c|}{0.71}   & \multicolumn{1}{c|}{0.71} & 0.94   & \multicolumn{1}{c|}{0.59}  & \multicolumn{1}{c|}{0.51}   & \multicolumn{1}{c|}{0.55} & 0.60   \\ 
Flink         & \multicolumn{1}{c|}{0.94}  & \multicolumn{1}{c|}{0.74}   & \multicolumn{1}{c|}{0.77} & 0.95   & \multicolumn{1}{c|}{0.62}  & \multicolumn{1}{c|}{0.60}   & \multicolumn{1}{c|}{0.62} & 0.71   \\ 
HBase         & \multicolumn{1}{c|}{0.91}  & \multicolumn{1}{c|}{0.77}   & \multicolumn{1}{c|}{0.80} & 0.92   & \multicolumn{1}{c|}{0.59}  & \multicolumn{1}{c|}{0.61}   & \multicolumn{1}{c|}{0.64} & 0.63   \\ 
JMeter        & \multicolumn{1}{c|}{0.92}  & \multicolumn{1}{c|}{0.73}   & \multicolumn{1}{c|}{0.74} & 0.95   & \multicolumn{1}{c|}{0.59}  & \multicolumn{1}{c|}{0.55}   & \multicolumn{1}{c|}{0.53} & 0.68   \\ 
Kafka         & \multicolumn{1}{c|}{0.93}  & \multicolumn{1}{c|}{0.68}   & \multicolumn{1}{c|}{0.70} & 0.98   & \multicolumn{1}{c|}{0.58}  & \multicolumn{1}{c|}{0.51}   & \multicolumn{1}{c|}{0.51} & 0.69   \\ 
Karaf         & \multicolumn{1}{c|}{0.93}  & \multicolumn{1}{c|}{0.73}   & \multicolumn{1}{c|}{0.79} & 0.94   & \multicolumn{1}{c|}{0.63}  & \multicolumn{1}{c|}{0.57}   & \multicolumn{1}{c|}{0.58} & 0.64   \\ 
Wicket        & \multicolumn{1}{c|}{0.94}  & \multicolumn{1}{c|}{0.74}   & \multicolumn{1}{c|}{0.75} & 0.95   & \multicolumn{1}{c|}{0.75}  & \multicolumn{1}{c|}{0.56}   & \multicolumn{1}{c|}{0.59} & 0.78   \\ 
Zookeeper     & \multicolumn{1}{c|}{0.92}  & \multicolumn{1}{c|}{0.68}   & \multicolumn{1}{c|}{0.74} & 0.94   & \multicolumn{1}{c|}{0.59}  & \multicolumn{1}{c|}{0.50}   & \multicolumn{1}{c|}{0.57} & 0.62   \\ \hline
Average       & \multicolumn{1}{c|}{0.93}  & \multicolumn{1}{c|}{0.74}   & \multicolumn{1}{c|}{0.75} & 0.95   & \multicolumn{1}{c|}{0.62}  & \multicolumn{1}{c|}{0.56}   & \multicolumn{1}{c|}{0.58} & 0.67   \\ \hline
\end{tabular}%
}
\end{table}
Next, we compare QuLog-8 against QuLog*. The results on the two evaluation criteria show that QuLog* outperforms QuLog-8 by 1-9\% on accuracy and 1-5\% on AUC for different systems. These results indicate the existence of shared system-agnostic properties of the static text and the log levels, independent of the software systems examined in the preliminary study. The instructions originate from different programming languages and publicly accessible software systems from GitHub, representing diverse developers writing styles. Therefore, by their leveraging, QuLog* learns a wide range of characteristics of the static text concerning the log levels (e.g., large vocabulary used in similar event descriptions). The good performance across different systems and the system-agnostic training of QuLog* suggest that QuLog is suitable for an automatic assessment of the quality of the log instructions, represented by their correct log level assignment. Examples of when QuLog is outperforming the baselines can be found in Appendix~\ref{examp}.
\subsubsection{Log Level Problem Instances} 
\begin{table}[!h]
\caption{Log level misclassification contingency table (the averaging is done over nine software systems given in \tablename~\ref{tab:deeplv})}
\label{tab:errordist}
\begin{tabular}{c|c|c|c}
\hline
True/Predicted & Info   & Warning & Error  \\ \hline
Info           & -      & 21.1\%  & 16.1\% \\ 
Warning        & 10.7\% & -       & 40.3\% \\ 
Error          & 4.3\%  & 19.3\%  & -      \\ \hline
\end{tabular}
\end{table}
The previous experiment shows that QuLog performs better than the baselines on log level assignments. However, the results between 0.60-0.78 on accuracy across different systems, although good, indicate that there are incorrect assignments. The misclassifications can impair the practical usability of QuLog. Therefore, we study the misclassification types. Based on the observations, we identified instances of the log level assignment problem having improved results facilitating the practical applicability of QuLog.  To study QuLog's misclassification types, we calculated the misclassification contingency table. It shows the percentage of misclassification prediction rates for the three classes. \tablename~\ref{tab:errordist} gives the contingency table. It is seen that some class pairs have a low misclassification rate (e.g., true "error" predicted as "info" is 4.3\%), however for others, it is significantly high (e.g., true "warning" predicted as "error" is 40.3\%). To understand the potential reasons, we examined the n-gram frequency shared between the different log levels similar to the preliminary study (Section ~\ref{esllq}). We find that n-grams shared between the log level pairs "error-warning" is 14.2\%, and it is higher compared to "error-info" (4.9\%) and "warning-info" (9.7\%). Relating it to the contingency matrix, we see that the class pairs with higher n-grams overlap have higher misclassification rates. We use this observation to construct three simplified instances of the log level quality assignment. Instead of predicting the three classes, we considered the prediction of two classes, namely "info-warning" (IW), "info-error" (IE) and "error-warning" (EW). The examination of individual class pairs has practical relevance because different stakeholders have different expectations from logs. For example, the operators usually examine the log levels "error" and "warning". Therefore, misclassifying an error event as "info" (e.g., Jira issue HDFS-4048) can hide important events from operators, increasing the maintenance costs. 

\textit{Experiment design.} We considered QuLog* log level assignment approach because it is system-agnostic. To train QuLog* on the three two-class problems, we modified the output layer to have two classes instead of three. The experiment is designed as follows. We start with the 100 software systems collected during the data collection procedure. We randomly sampled 60\% of the repositories for training, 20\% for validation and 20\% for evaluation. To reduce the variance of the results due to the random repositories selection, we repeated the sampling procedure 30 times and reported the average results alongside the standard deviations. To assess the correctness of the decisions, we used F$_1$, precision and recall, instead of accuracy because they are exposing the imbalances of the class distributions better than accuracy. We used the same baselines as in the previous experiment trained with the same data as QuLog*.
\begin{table}[!h]
\caption{Performance scores on the task of log level assignment. The best results per scenario are bolded.}
\label{tab1:loglevelprediction}
\resizebox{\columnwidth}{!}{%
\begin{tabular}{c|l|c|c|c}
\hline
Scores                     & Scenario & QuLog     & DeepLV     & BERT\_SVM \\ \hline
\multirow{4}{*}{F$_1$}        & IE       & \textbf{0.88±0.03} & 0.82±0.02  & 0.87±0.04 \\ \cline{2-5} 
                           & IWE      & \textbf{0.73±0.03} & 0.67±0.03  & \textbf{0.73±0.04} \\ \cline{2-5} 
                           & IW       & \textbf{0.68±0.06} & 0.61±0.08  & 0.64±0.05 \\ \cline{2-5} 
                           & WE       & \textbf{0.61±0.04} & 0.56±0.06  & 0.54±0.05 \\ \hline
\multirow{4}{*}{Precision} & IE       & 0.88±0.02 & 0.79±0.04  & \textbf{0.92±0.01} \\ \cline{2-5} 
                           & IWE      & 0.72±0.03 & 0.66±0.03 & \textbf{0.74±0.05} \\ \cline{2-5} 
                           & IW       & \textbf{0.75±0.04} & 0.72±0.06 & 0.58±0.05 \\ \cline{2-5} 
                           & WE       & \textbf{0.69±0.09} & 0.59±0.08 & 0.51±0.07 \\ \hline
\multirow{4}{*}{Recall}    & IE       & \textbf{0.89±0.05} & 0.86±0.06  & 0.84±0.06 \\ \cline{2-5} 
                           & IWE      & \textbf{0.73±0.03} & 0.68±0.03 &  \textbf{0.73±0.04 }\\ \cline{2-5}
                           & IW       & 0.62±0.08 & 0.54±0.1  & \textbf{0.72±0.09} \\ \cline{2-5} 
                           & WE       & 0.56±0.07 & 0.54±0.08 &  \textbf{0.59±0.08} \\ \hline
\end{tabular}%
}
\end{table}

\textit{Results and discussion}. \tablename~\ref{tab1:loglevelprediction} enlists the performance scores for the four problem instances of log level quality assessment. Comparing the absolute values for the scores across the four scenarios reveals that the IE problem achieves the highest values on F$_1$ score (average of 0.88), i.e., trades-off the precision (0.88) and recall (0.89) quite well. Therefore, this model is very reliable for correctly assessing the "info" and "error" log instructions. The good performance is attributed to the observed differences in the vocabulary between the "error" and "info" log instructions (i.e., a 4.9\% n-gram overlap). Therefore, this model won't overwhelm developers with many incorrect predictions. On the IW and EW problem instances, although QuLog does not perform as good, still outperform the baselines when different software systems are considered.

\subsection{Linguistic Quality Assessment Evaluation}
\subsubsection{Experimental Design} To evaluate the sufficiency in the linguistic structure of the static text, we used the data from the preliminary study as given in Section~\ref{lqa}. We trained QuLog on the linguistic representations from the eight systems and evaluated the remaining one. Notably,  we identify log instructions with insufficient linguistic structure in four of the tested systems: Cassandra, HBase, Kafka and Zookeeper, and we report the results for them. 
As baselines, we considered two popular binary text classification methods, i.e., SVM and Random Forest (RF)~\cite{Breiman2001}, trained on the general-purpose representation of the linguistic categories~(BERT)~\cite{devlin2018bert}. We train QuLog's linguistic quality assessment part with the same values of the hyperparameters as for the log level quality assessment, with setting the batch size to 64. As evaluation criteria, we used F$_1$ and specificity. Additional evaluation against rule-based approaches can be found in Appendix~\ref{rule-basedapproach}.
\begin{table}[!h]
\caption{Sufficient linguistic structure assessment (performance evaluation)}
\label{tab:lqares}
\resizebox{\columnwidth}{!}{%
\begin{tabular}{c|ccc|ccc}
\hline
          & \multicolumn{3}{c|}{F$_1$}                                       & \multicolumn{3}{c}{Specificity}         \\ \hline
System    & \multicolumn{1}{c|}{QuLog} & \multicolumn{1}{c|}{SVM}  & RF   & \multicolumn{1}{c|}{QuLog} & \multicolumn{1}{c|}{SVM}  & RF   \\ \hline
Cassandra & \multicolumn{1}{c|}{1.00}  & \multicolumn{1}{c|}{0.99} & 0.99 & \multicolumn{1}{c|}{1.00}  & \multicolumn{1}{c|}{1.00} & 0.96 \\ 
HBase     & \multicolumn{1}{c|}{0.96}  & \multicolumn{1}{c|}{0.96} & 0.97 & \multicolumn{1}{c|}{0.97}  & \multicolumn{1}{c|}{0.94} & 0.92 \\ 
Kafka     & \multicolumn{1}{c|}{0.99}  & \multicolumn{1}{c|}{0.98} & 0.92 & \multicolumn{1}{c|}{1.00}  & \multicolumn{1}{c|}{1.00} & 0.74 \\ 
Zookeeper & \multicolumn{1}{c|}{0.99}  & \multicolumn{1}{c|}{0.99} & 0.98 & \multicolumn{1}{c|}{1.00}  & \multicolumn{1}{c|}{0.98} & 0.94 \\ \hline
Average   & \multicolumn{1}{c|}{0.98}  & \multicolumn{1}{c|}{0.97} & 0.96 & \multicolumn{1}{c|}{0.99}  & \multicolumn{1}{c|}{0.98} & 0.89 \\ \hline
\end{tabular}%
}
\end{table}
\sloppy{\subsubsection{Results and discussion} \tablename~\ref{tab:lqares} enlists the evaluation results. It is seen that QuLog achieves a high average F$_1$ score of 0.98 while outperforming the baselines by slight margins. The good performance of the three methods is attributed to the discriminative linguistic features between the two classes. For example,  the HBase's log instruction "failed parse", from the class \textit{hadoop.hbase.zookeeper.ZKListener}, has a linguistic structure "verb noun". Notably, it does not contain information to which the parsing failure refers (i.e., lacks sufficient linguistic structure). As a comparison, in another log instruction "failed parse data for znode *" within the same class of HBase, the linguistic structure "verb noun" has four additional linguistic properties, i.e., it has the form "verb noun noun apposition noun parameter". This additional linguistic structure has two advantages. From a learning perspective, the richer linguistic structure is useful for discriminating between the classes. From a comprehension perspective, it encodes verbose information on the type of failed parsing. The better performance of QuLog against the baselines can be attributed to its ability to extract a better representation of the linguistic structure. QuLog exploits log specific concepts, while the general-purpose language models are trained on datasets from general literature, which may average-out log specific properties.}

The results on \textit{specificity} are high for both QuLog and SVM while being a bit lower for RF. Since specificity evaluates methods' performance in the correct prediction for the insufficient class (true negative class), the results show that QuLog can correctly identify the instructions with an insufficient linguistic structure. By combing these results with the high performance on F$_1$ (as a trade-off between incorrect sufficient predictions), we conclude that QuLog detects the linguistically insufficient instructions without compromising the performance on the sufficient class. The high values for the two scores show that QuLog is useful in automatically assessing the sufficient linguistic structure in a system-agnostic manner.  By pointing out the log instructions that may benefit from enriching the static text, QuLog improves the comprehensibility of the log instructions, ultimately improving the logging quality.
\subsection{Prediction Explainer Evaluation}
\textit{Experiment design.} To evaluate the prediction explainer, we construct an artificial dataset as in the following. We start by randomly sampling 100 static texts of the instructions with a correct log level prediction of an already trained model (i.e., QuLog* for log level assignment). Each static text is manually investigated by two developers and modified such that a manually selected word of the static text is replaced with its antonym. This changes the event description creating an opposite event. For example, we start with the original static text "Connection established" with the original log level "info". We change the token "established" into its antonym "refused", obtaining a modified static text, i.e., "Connection refused" and modified word "refused". Since the modified token is an antonym,  we change the original log level ("info") into the modified log level ("error"). Therefore, for each static text we obtain a tuple of five elements -- 1) original static text, 2) modified static text, 3) modified word, 4) original level, and 5) modified level. From the initial 100, the two annotators initially agreed on 42 changes. In a subsequent discussion, the number was increased to 65 instructions used in the evaluation. The original and modified static texts are given to the \textit{prediction explainer} that generates the ordered token list of importance scores. The modified token is used as ground truth. We check how many tokens should the developer examine before finding the modified token, and we measure it by the $error@k$ performance score. We considered two log level models, the two-class IE, due to its high performance and the three-class log level assignment IWE. As a baseline, we consider suggesting a randomly chosen token as the most relevant. 

\begin{figure}[!b]
\centering
\includegraphics[width=\columnwidth]{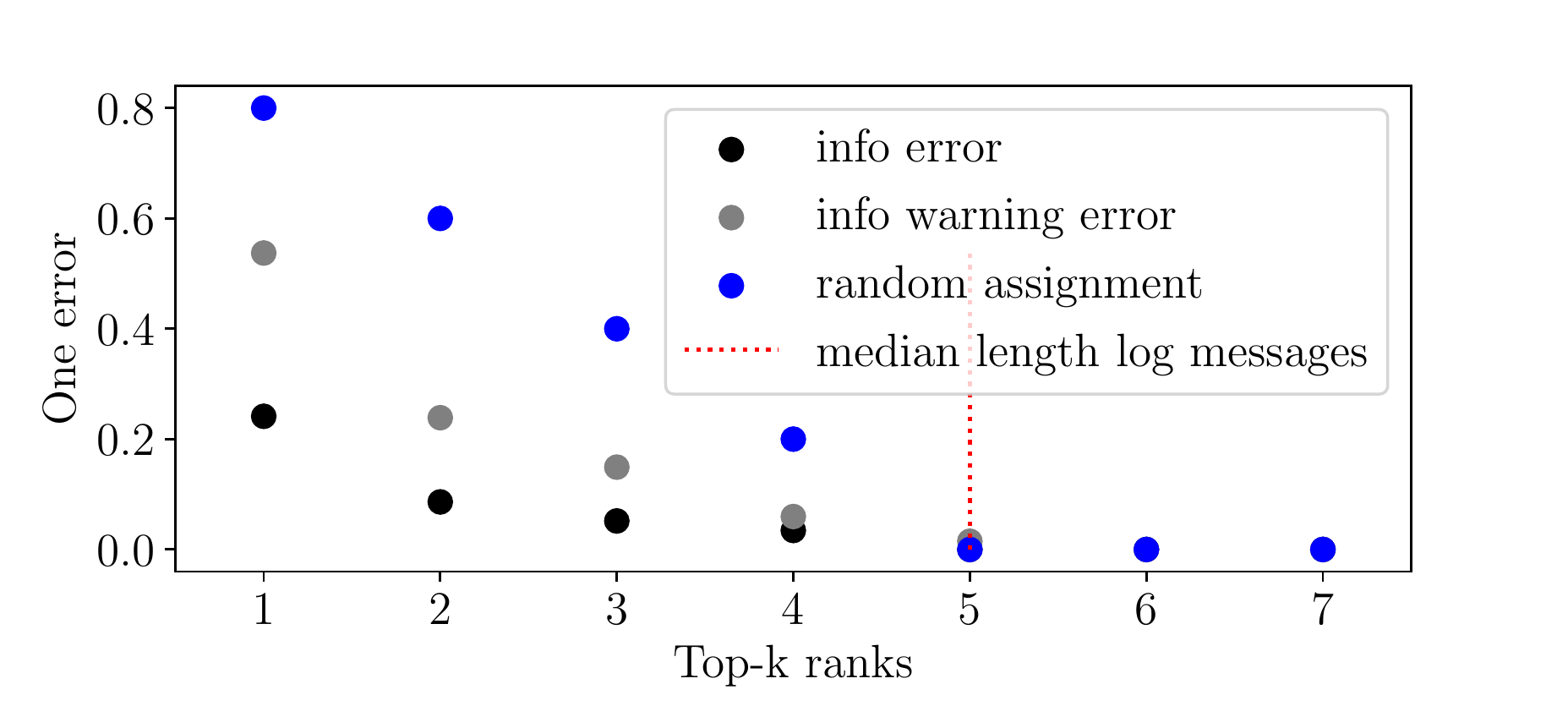}
\caption{Evaluation of the explanation module}
\label{fig:interpretability:oneerror}
\end{figure}

\textit{Results and discussion} \figurename~\ref{fig:interpretability:oneerror} depicts the experimental results. It is seen that the prediction explanation module has a low error on correct word suggestions (the error@1 is 0.25) for the IE model. The prediction explanation model for the IWE model is a bit higher (the error@1 is 0.52), however, both explainers show better performance than the considered baseline. The observed discrepancy between the prediction explanations of the IE and IWE is due to the better performance of the IE model (average F$_1$ score 0.88) as opposed to the IWE model (average F$_1$ score of 0.73). It indicates that a better performing model learns discriminative features better. By considering $k$ relevant tokens (i.e., a developer examines the k highest-ranked tokens), the three explanation models reduce the error, with IW and IWE having sharp decreases, achieving 0.05 and 0.23 on error@2 correspondingly. The low value of the one error shows that QuLog's log level prediction explainer correctly explains the predictions. 
Therefore, the prediction explainer gives valuable suggestions on static text updates to improve quality. 

\subsection{Use Cases}
We further applied QuLog on two internal systems. For log level assignment, we considered QuLog* IE model. \tablename~\ref{usecases} summarizes the results. For System 1, on the task of log level assignment, QuLog agrees in 52/63 cases with the original log levels and 56/63 on sufficient linguistic structure assessment. Developers examined the 11 disagreements for log level and the seven disagreements on sufficient linguistic structure. They decided to change 5/11 log levels and augmented 4/7 with additional linguistic structures. The remaining linguistic suggestions were considered as "unimportant". Similarly, for System 2, QuLog agreed in 120/138 cases on log level, with 8/18 log levels being changed. On the sufficient linguistic evaluation, QuLog identified five instructions with insufficient static text, three of which were accepted. These two examples showcase how QuLog's automatically assess the log instructions, giving useful suggestions for improving the logging quality.
\begin{table}[!h]
\caption{Use Case Study Results}
\label{usecases}
\resizebox{\columnwidth}{!}{%
\begin{tabular}{l|c|c|c} 
\cline{2-4}
\multicolumn{1}{l|}{} & Log Instructions & \begin{tabular}[c]{@{}c@{}}Log Level\\Recommendations\end{tabular} & \begin{tabular}[c]{@{}c@{}}Linguistic Structure\\Recommendations\end{tabular}  \\ 
\hline
System 1              & 63               & 52 (5/11)                                                          & 56 (4/7)                                                                       \\ 

System 2              & 138              & 120 (8/18)~                                                        & \textcolor[rgb]{0.125,0.129,0.141}{133 (3/5)}                                  \\
\hline
\end{tabular}%
}
\end{table}
\subsection{Threats to Validity}~\label{sec5}
The key threats to the validity of this study related are to the included datasets and the implementation details. We chose vetted systems for the preliminary study following related works~\cite{DeepLV}. We further complemented it with another public dataset collection to mitigate data selection artefacts. The datasets used for the sufficient linguistic structure and prediction explainer might be biased by the human annotation procedure. Therefore, we considered two annotators to construct each of them. A third-party evaluation may help to further mitigate the biases from the annotation.

\section{Related work}~\label{sec6}
\textbf{Logging practices} We discussed the studies on quality properties in Section~\ref{sec21}. Alongside the quality properties, several literature studies are examining diverse logging practices. In one of the first studies, Yuan et al.~\cite{Yuan2012} quantify the log pervasiveness and the benefit of software logging in C/C++ systems while proposing proactive logging strategies. They found that developers spent significant efforts modifying the log levels, static texts, and the parameters of log instructions but do not change their locations often. Similar observations are made for Java~\cite{Chen2017, Kabina2016, Sheng2014, Chen2019, taoxie} and Android software systems~\cite{Zeng2019}. These studies augment the aforenamed by introducing new research questions like studying log bug resolution time~\cite{Chen2017}, log instruction update types~\cite{Chen2019} and evolution of logging configuration~\cite{taoxie}. For example, Kabina et al.~\cite{Kabina2016} identified that 20-45\% of the log instructions change through system lifetime. The importance of logging practices is widely recognized in the industry,  seen by several logging practice studies of industrial systems~\cite{microsoftTelemetry, whereToLogStudy2021, Pecchia2015}. In a field study, Li et al.~\cite{Hassan2020} demonstrates the different costs of logging from developers and research perspectives. The similarities of the static text concerning the different log levels and the similarities in the conclusions regarding the various logging practices in different programming languages motivate our work on software systems cross-examination when evaluating log instruction quality.

\textbf{Automatic Logging Enhancement.} There are several methods that support the automatic enhancement of log instructions~\cite{zhenhao2020, SoftwareLoggingTopicModels, Li2021, LogAdvisor, ErrLog}. Based on the enhancement type, we distinguish two groups of methods, i.e., methods addressing the log instructions placement problem (where-to-log)~\cite{whereToLogCotroneo, Fu2014, whereToLogStudy2021, SoftwareLoggingTopicModels, Li2020}, and methods addressing the choice of relevant logging information (what-to-log)~\cite{LogEnchancer, ErrLog, codecomp, Li2021}. The latter can further be separated into three groups: 1) log message generation~\cite{PinjiaHe2018}, 2) relevant variables placement~\cite{20Log}, and 3) log level suggestion~\cite{DeepLV, Ann2019}. Different from previous work, we utilize shared language properties between diverse software systems to develop an automatic system-agnostic approach for log instruction quality assessment, defined as 1) log level assignment and 2) sufficient linguistic structure assessments.

\section{Conclusion}~\label{sec7}
Writing log instruction with sufficient quality is challenging due to the absence of complete logging guidelines and developers incomplete understanding of system complexity. 
In this work, we address the problem of automating log quality assessment. We first do a preliminary study on nine software systems to study the quality properties of the log instructions. The results of our study identified 1) log level assignment and 2) sufficient linguistic structure assessments as two quality properties identifiable solely by the static text of the log instruction. Based on our observations, we propose a deep learning-based approach for automatic log instruction quality assessment on a given target system. Our approach uses the static text and its linguistic structure representation to evaluate the two properties. In addition, we adopt an approach from explainable AI to reason about model predictions and give suggestions for potential improvements of the instruction. Our approach outperforms the considered baselines, achieving high accuracy for log level assignment (0.88) and a high F$_1$ score for sufficient linguistic structure assessment (0.99). The results highlight future research opportunities in using cross-systems log instructions not just in automatically assessing log instruction quality, but also to automatically enhance them (i.e.,  automatic log instruction writing).

\bibliographystyle{ACM-Reference-Format}
\bibliography{sample-base}


\begin{thebibliography}{53}


\ifx \showCODEN    \undefined \def \showCODEN     #1{\unskip}     \fi
\ifx \showDOI      \undefined \def \showDOI       #1{#1}\fi
\ifx \showISBNx    \undefined \def \showISBNx     #1{\unskip}     \fi
\ifx \showISBNxiii \undefined \def \showISBNxiii  #1{\unskip}     \fi
\ifx \showISSN     \undefined \def \showISSN      #1{\unskip}     \fi
\ifx \showLCCN     \undefined \def \showLCCN      #1{\unskip}     \fi
\ifx \shownote     \undefined \def \shownote      #1{#1}          \fi
\ifx \showarticletitle \undefined \def \showarticletitle #1{#1}   \fi
\ifx \showURL      \undefined \def \showURL       {\relax}        \fi
\providecommand\bibfield[2]{#2}
\providecommand\bibinfo[2]{#2}
\providecommand\natexlab[1]{#1}
\providecommand\showeprint[2][]{arXiv:#2}

\bibitem[\protect\citeauthoryear{Barik, DeLine, Drucker, and Fisher}{Barik
  et~al\mbox{.}}{2016}]%
        {microsoftTelemetry}
\bibfield{author}{\bibinfo{person}{Titus Barik}, \bibinfo{person}{Robert
  DeLine}, \bibinfo{person}{Steven Drucker}, {and} \bibinfo{person}{Danyel
  Fisher}.} \bibinfo{year}{2016}\natexlab{}.
\newblock \showarticletitle{The Bones of the System: A Case Study of Logging
  and Telemetry at Microsoft}. In \bibinfo{booktitle}{\emph{2016 IEEE/ACM 38th
  International Conference on Software Engineering Companion (ICSE-C)}}.
  \bibinfo{publisher}{Association for Computing Machinery},
  \bibinfo{address}{New York, NY, USA}, \bibinfo{pages}{92--101}.
\newblock


\bibitem[\protect\citeauthoryear{Bogatinovski}{Bogatinovski}{2021}]%
        {githubrepoVisible}
\bibfield{author}{\bibinfo{person}{Jasmin Bogatinovski}.}
  \bibinfo{year}{2021}\natexlab{}.
\newblock \bibinfo{title}{QuLog: Github repo link}.
\newblock   (\bibinfo{year}{2021}).
\newblock
\urldef\tempurl%
\url{https://github.com/qulog/QuLog}
\showURL{%
Retrieved January 02, 2022 from \tempurl}


\bibitem[\protect\citeauthoryear{Breiman}{Breiman}{2001}]%
        {Breiman2001}
\bibfield{author}{\bibinfo{person}{Leo Breiman}.}
  \bibinfo{year}{2001}\natexlab{}.
\newblock \showarticletitle{Random Forests}.
\newblock \bibinfo{journal}{\emph{Machine Learning}} \bibinfo{volume}{45},
  \bibinfo{number}{1} (\bibinfo{year}{2001}), \bibinfo{pages}{5--32}.
\newblock


\bibitem[\protect\citeauthoryear{Brown, Mann, Ryder, Subbiah, Kaplan, Dhariwal,
  Neelakantan, Shyam, Sastry, Askell, Agarwal, Herbert-Voss, Krueger, Henighan,
  Child, Ramesh, Ziegler, Wu, Winter, Hesse, Chen, Sigler, Litwin, Gray, Chess,
  Clark, Berner, McCandlish, Radford, Sutskever, and Amodei}{Brown
  et~al\mbox{.}}{2020}]%
        {GPT3}
\bibfield{author}{\bibinfo{person}{Tom Brown}, \bibinfo{person}{Benjamin Mann},
  \bibinfo{person}{Nick Ryder}, \bibinfo{person}{Melanie Subbiah},
  \bibinfo{person}{Jared~D Kaplan}, \bibinfo{person}{Prafulla Dhariwal},
  \bibinfo{person}{Arvind Neelakantan}, \bibinfo{person}{Pranav Shyam},
  \bibinfo{person}{Girish Sastry}, \bibinfo{person}{Amanda Askell},
  \bibinfo{person}{Sandhini Agarwal}, \bibinfo{person}{Ariel Herbert-Voss},
  \bibinfo{person}{Gretchen Krueger}, \bibinfo{person}{Tom Henighan},
  \bibinfo{person}{Rewon Child}, \bibinfo{person}{Aditya Ramesh},
  \bibinfo{person}{Daniel Ziegler}, \bibinfo{person}{Jeffrey Wu},
  \bibinfo{person}{Clemens Winter}, \bibinfo{person}{Chris Hesse},
  \bibinfo{person}{Mark Chen}, \bibinfo{person}{Eric Sigler},
  \bibinfo{person}{Mateusz Litwin}, \bibinfo{person}{Scott Gray},
  \bibinfo{person}{Benjamin Chess}, \bibinfo{person}{Jack Clark},
  \bibinfo{person}{Christopher Berner}, \bibinfo{person}{Sam McCandlish},
  \bibinfo{person}{Alec Radford}, \bibinfo{person}{Ilya Sutskever}, {and}
  \bibinfo{person}{Dario Amodei}.} \bibinfo{year}{2020}\natexlab{}.
\newblock \showarticletitle{Language Models are Few-Shot Learners}. In
  \bibinfo{booktitle}{\emph{Advances in Neural Information Processing
  Systems}}, Vol.~\bibinfo{volume}{33}. \bibinfo{publisher}{Curran Associates,
  Inc.}, \bibinfo{pages}{1877--1901}.
\newblock


\bibitem[\protect\citeauthoryear{C\^{a}ndido, Jan, Aniche, and van
  Deursen}{C\^{a}ndido et~al\mbox{.}}{2021}]%
        {whereToLogStudy2021}
\bibfield{author}{\bibinfo{person}{Jeanderson C\^{a}ndido},
  \bibinfo{person}{Haesen Jan}, \bibinfo{person}{Maurício Aniche}, {and}
  \bibinfo{person}{Arie van Deursen}.} \bibinfo{year}{2021}\natexlab{}.
\newblock \showarticletitle{An Exploratory Study of Log Placement
  Recommendation in an Enterprise System}. In \bibinfo{booktitle}{\emph{2021
  IEEE/ACM 18th International Conference on Mining Software Repositories
  (MSR)}}. \bibinfo{publisher}{IEEE Computer Society}, \bibinfo{address}{Los
  Alamitos, CA, USA}, \bibinfo{pages}{143--154}.
\newblock
\urldef\tempurl%
\url{https://doi.org/10.1109/MSR52588.2021.00027}
\showDOI{\tempurl}


\bibitem[\protect\citeauthoryear{Chen and (Jack)~Jiang}{Chen and
  (Jack)~Jiang}{2017}]%
        {Chen2017}
\bibfield{author}{\bibinfo{person}{Boyuan Chen} {and}
  \bibinfo{person}{Zhen~Ming (Jack)~Jiang}.} \bibinfo{year}{2017}\natexlab{}.
\newblock \showarticletitle{Characterizing logging practices in Java-based open
  source software projects -- a replication study in Apache Software
  Foundation}.
\newblock \bibinfo{journal}{\emph{Empirical Software Engineering}}
  \bibinfo{volume}{22} (\bibinfo{year}{2017}), \bibinfo{pages}{330--374}.
\newblock


\bibitem[\protect\citeauthoryear{Chen and Jiang}{Chen and Jiang}{2019}]%
        {Chen2019}
\bibfield{author}{\bibinfo{person}{Boyuan Chen} {and}
  \bibinfo{person}{Zhen~Ming Jiang}.} \bibinfo{year}{2019}\natexlab{}.
\newblock \showarticletitle{Extracting and Studying the Logging-Code-Issue-
  Introducing Changes in Java-Based Large-Scale Open Source Software Systems}.
\newblock \bibinfo{journal}{\emph{Empirical Softw. Engg.}}
  \bibinfo{volume}{24} (\bibinfo{year}{2019}), \bibinfo{pages}{2285–2322}.
\newblock


\bibitem[\protect\citeauthoryear{Chen, Thomas, and Hassan}{Chen
  et~al\mbox{.}}{2016}]%
        {28}
\bibfield{author}{\bibinfo{person}{Tse-Hsun Chen}, \bibinfo{person}{Stephen~W.
  Thomas}, {and} \bibinfo{person}{Ahmed~E. Hassan}.}
  \bibinfo{year}{2016}\natexlab{}.
\newblock \showarticletitle{A Survey on the Use of Topic Models When Mining
  Software Repositories}.
\newblock \bibinfo{journal}{\emph{Empirical Softw. Engg.}}
  \bibinfo{volume}{21} (\bibinfo{year}{2016}), \bibinfo{pages}{1843–1919}.
\newblock


\bibitem[\protect\citeauthoryear{Cinque, Cotroneo, and Pecchia}{Cinque
  et~al\mbox{.}}{2013}]%
        {whereToLogCotroneo}
\bibfield{author}{\bibinfo{person}{Marcello Cinque}, \bibinfo{person}{Domenico
  Cotroneo}, {and} \bibinfo{person}{Antonio Pecchia}.}
  \bibinfo{year}{2013}\natexlab{}.
\newblock \showarticletitle{Event Logs for the Analysis of Software Failures: A
  Rule-Based Approach}.
\newblock \bibinfo{journal}{\emph{IEEE Transactions on Software Engineering}}
  \bibinfo{volume}{39} (\bibinfo{year}{2013}), \bibinfo{pages}{806--821}.
\newblock


\bibitem[\protect\citeauthoryear{Cortes and Vapnik}{Cortes and Vapnik}{1995}]%
        {Vapnik1995}
\bibfield{author}{\bibinfo{person}{Corinna Cortes} {and}
  \bibinfo{person}{Vladimir Vapnik}.} \bibinfo{year}{1995}\natexlab{}.
\newblock \showarticletitle{Support-vector networks}.
\newblock \bibinfo{journal}{\emph{Machine Learning}}  \bibinfo{volume}{20}
  (\bibinfo{year}{1995}), \bibinfo{pages}{273--297}.
\newblock


\bibitem[\protect\citeauthoryear{Cover and Thomas}{Cover and Thomas}{2006}]%
        {informationtheory}
\bibfield{author}{\bibinfo{person}{Thomas~M. Cover} {and}
  \bibinfo{person}{Joy~A. Thomas}.} \bibinfo{year}{2006}\natexlab{}.
\newblock \bibinfo{booktitle}{\emph{Elements of Information Theory (Wiley
  Series in Telecommunications and Signal Processing)}}.
\newblock \bibinfo{publisher}{Wiley-Interscience}, \bibinfo{address}{USA}.
\newblock
\showISBNx{0471241954}


\bibitem[\protect\citeauthoryear{Devlin, Chang, Lee, and Toutanova}{Devlin
  et~al\mbox{.}}{2019}]%
        {devlin2018bert}
\bibfield{author}{\bibinfo{person}{Jacob Devlin}, \bibinfo{person}{Ming-Wei
  Chang}, \bibinfo{person}{Kenton Lee}, {and} \bibinfo{person}{Kristina
  Toutanova}.} \bibinfo{year}{2019}\natexlab{}.
\newblock \bibinfo{title}{Bert: Pre-training of deep bidirectional transformers
  for language understanding}.
\newblock   (\bibinfo{year}{2019}), \bibinfo{numpages}{4171–-4186}~pages.
\newblock
\urldef\tempurl%
\url{https://doi.org/10.18653/v1/N19-1423}
\showDOI{\tempurl}


\bibitem[\protect\citeauthoryear{Ding, Zhou, Lou, Zhang, Lin, Fu, Zhang, and
  Xie}{Ding et~al\mbox{.}}{2015}]%
        {log2}
\bibfield{author}{\bibinfo{person}{Rui Ding}, \bibinfo{person}{Hucheng Zhou},
  \bibinfo{person}{Jian-Guang Lou}, \bibinfo{person}{Hongyu Zhang},
  \bibinfo{person}{Qingwei Lin}, \bibinfo{person}{Qiang Fu},
  \bibinfo{person}{Dongmei Zhang}, {and} \bibinfo{person}{Tao Xie}.}
  \bibinfo{year}{2015}\natexlab{}.
\newblock \showarticletitle{Log2: A Cost-Aware Logging Mechanism for
  Performance Diagnosis}. In \bibinfo{booktitle}{\emph{Proceedings of the 2015
  USENIX Conference on Usenix Annual Technical Conference}}.
  \bibinfo{publisher}{USENIX Association}, \bibinfo{address}{USA},
  \bibinfo{pages}{139–150}.
\newblock


\bibitem[\protect\citeauthoryear{Finegan}{Finegan}{2014}]%
        {Finegan2014}
\bibfield{author}{\bibinfo{person}{Edward Finegan}.}
  \bibinfo{year}{2014}\natexlab{}.
\newblock \bibinfo{booktitle}{\emph{Language: Its structure and use}
  (\bibinfo{edition}{7} ed.)}.
\newblock \bibinfo{publisher}{Cengage Learning}, \bibinfo{address}{Florence,
  AL}. 289 pages.
\newblock


\bibitem[\protect\citeauthoryear{Foundation}{Foundation}{2022}]%
        {Log4J}
\bibfield{author}{\bibinfo{person}{The Apache~Software Foundation}.}
  \bibinfo{year}{2022}\natexlab{}.
\newblock \bibinfo{title}{Logging Service Project}.
\newblock   (\bibinfo{year}{2022}).
\newblock
\urldef\tempurl%
\url{https://logging.apache.org/}
\showURL{%
Retrieved January 2, 2022 from \tempurl}


\bibitem[\protect\citeauthoryear{Fu, Zhu, Hu, Lou, Ding, Lin, Zhang, and
  Xie}{Fu et~al\mbox{.}}{2014}]%
        {Fu2014}
\bibfield{author}{\bibinfo{person}{Qiang Fu}, \bibinfo{person}{Jieming Zhu},
  \bibinfo{person}{Wenlu Hu}, \bibinfo{person}{Jian-Guang Lou},
  \bibinfo{person}{Rui Ding}, \bibinfo{person}{Qingwei Lin},
  \bibinfo{person}{Dongmei Zhang}, {and} \bibinfo{person}{Tao Xie}.}
  \bibinfo{year}{2014}\natexlab{}.
\newblock \showarticletitle{Where Do Developers Log? An Empirical Study on
  Logging Practices in Industry}. In \bibinfo{booktitle}{\emph{Companion
  Proceedings of the 36th International Conference on Software Engineering}}.
  \bibinfo{publisher}{Association for Computing Machinery},
  \bibinfo{address}{New York, NY, USA}, \bibinfo{pages}{24–33}.
\newblock


\bibitem[\protect\citeauthoryear{Gibaja and Ventura}{Gibaja and
  Ventura}{2015}]%
        {Gibaja2014}
\bibfield{author}{\bibinfo{person}{Eva Gibaja} {and}
  \bibinfo{person}{Sebasti\'{a}n Ventura}.} \bibinfo{year}{2015}\natexlab{}.
\newblock \showarticletitle{A Tutorial on Multilabel Learning}.
\newblock \bibinfo{journal}{\emph{Comput. Surveys}} \bibinfo{volume}{47},
  \bibinfo{number}{3} (\bibinfo{year}{2015}), \bibinfo{pages}{52:1--52:38}.
\newblock


\bibitem[\protect\citeauthoryear{Han, Jie, Wenchang, Jianwei, Bin, and Bo}{Han
  et~al\mbox{.}}{2019}]%
        {Ann2019}
\bibfield{author}{\bibinfo{person}{Anu Han}, \bibinfo{person}{Chen Jie},
  \bibinfo{person}{Shi Wenchang}, \bibinfo{person}{Hou Jianwei},
  \bibinfo{person}{Liang Bin}, {and} \bibinfo{person}{Qin Bo}.}
  \bibinfo{year}{2019}\natexlab{}.
\newblock \showarticletitle{An Approach to Recommendation of Verbosity Log
  Levels Based on Logging Intention}. In \bibinfo{booktitle}{\emph{2019 IEEE
  International Conference on Software Maintenance and Evolution (ICSME)}}.
  \bibinfo{publisher}{IEEE}, \bibinfo{address}{New York, USA},
  \bibinfo{pages}{125--134}.
\newblock


\bibitem[\protect\citeauthoryear{Hand}{Hand}{2001}]%
        {auc}
\bibfield{author}{\bibinfo{person}{Robert~J. Hand, David J.and~Till}.}
  \bibinfo{year}{2001}\natexlab{}.
\newblock \showarticletitle{A Simple Generalisation of the Area Under the ROC
  Curve for Multiple Class Classification Problems}.
\newblock \bibinfo{journal}{\emph{Machine Learning}}  \bibinfo{volume}{45}
  (\bibinfo{year}{2001}), \bibinfo{pages}{171--186}.
\newblock


\bibitem[\protect\citeauthoryear{Hassan}{Hassan}{2009}]%
        {ShanonEntropy}
\bibfield{author}{\bibinfo{person}{Ahmed~E. Hassan}.}
  \bibinfo{year}{2009}\natexlab{}.
\newblock \showarticletitle{Predicting faults using the complexity of code
  changes}. In \bibinfo{booktitle}{\emph{2009 IEEE 31st International
  Conference on Software Engineering}}. \bibinfo{publisher}{IEEE Computer
  Society}, \bibinfo{address}{USA}, \bibinfo{pages}{78--88}.
\newblock
\urldef\tempurl%
\url{https://doi.org/10.1109/ICSE.2009.5070510}
\showDOI{\tempurl}


\bibitem[\protect\citeauthoryear{Hassani, Shang, Shihab, and Tsantalis}{Hassani
  et~al\mbox{.}}{2018}]%
        {WeviShen2018}
\bibfield{author}{\bibinfo{person}{Mehran Hassani}, \bibinfo{person}{Weiyi
  Shang}, \bibinfo{person}{Emad Shihab}, {and} \bibinfo{person}{Nikolaos
  Tsantalis}.} \bibinfo{year}{2018}\natexlab{}.
\newblock \showarticletitle{Studying and Detecting Log-Related Issues}.
\newblock \bibinfo{journal}{\emph{Empirical Softw. Engg.}}
  \bibinfo{volume}{23} (\bibinfo{year}{2018}), \bibinfo{pages}{3248–3280}.
\newblock
\urldef\tempurl%
\url{https://doi.org/10.1007/s10664-018-9603-z}
\showDOI{\tempurl}


\bibitem[\protect\citeauthoryear{He, Chen, He, and Lyu}{He
  et~al\mbox{.}}{2018}]%
        {PinjiaHe2018}
\bibfield{author}{\bibinfo{person}{Pinjia He}, \bibinfo{person}{Zhuangbin
  Chen}, \bibinfo{person}{Shilin He}, {and} \bibinfo{person}{Michael~R. Lyu}.}
  \bibinfo{year}{2018}\natexlab{}.
\newblock \showarticletitle{Characterizing the Natural Language Descriptions in
  Software Logging Statements}. In \bibinfo{booktitle}{\emph{Proceedings of the
  33rd ACM/IEEE International Conference on Automated Software Engineering}}.
  \bibinfo{publisher}{Association for Computing Machinery},
  \bibinfo{address}{New York, NY, USA}, \bibinfo{pages}{178–189}.
\newblock


\bibitem[\protect\citeauthoryear{Honnibal, Montani, Van~Landeghem, and
  Boyd}{Honnibal et~al\mbox{.}}{2020}]%
        {spacy}
\bibfield{author}{\bibinfo{person}{Matthew Honnibal}, \bibinfo{person}{Ines
  Montani}, \bibinfo{person}{Sofie Van~Landeghem}, {and}
  \bibinfo{person}{Adriane Boyd}.} \bibinfo{year}{2020}\natexlab{}.
\newblock \bibinfo{title}{{spaCy: Industrial-strength Natural Language
  Processing in Python}}.
\newblock   (\bibinfo{year}{2020}).
\newblock
\urldef\tempurl%
\url{https://doi.org/10.5281/zenodo.1212303}
\showDOI{\tempurl}


\bibitem[\protect\citeauthoryear{Joachims}{Joachims}{2005}]%
        {Thorsten2005}
\bibfield{author}{\bibinfo{person}{Thorsten Joachims}.}
  \bibinfo{year}{2005}\natexlab{}.
\newblock \showarticletitle{A Support Vector Method for Multivariate
  Performance Measures}. In \bibinfo{booktitle}{\emph{Proceedings of the 22nd
  International Conference on Machine Learning}}.
  \bibinfo{publisher}{Association for Computing Machinery},
  \bibinfo{address}{New York, NY, USA}, \bibinfo{pages}{377–384}.
\newblock


\bibitem[\protect\citeauthoryear{Kabinna, Bezemer, Shang, Syer, and
  Hassan}{Kabinna et~al\mbox{.}}{2018}]%
        {Kabina2016}
\bibfield{author}{\bibinfo{person}{Suhas Kabinna}, \bibinfo{person}{Cor-Paul
  Bezemer}, \bibinfo{person}{Weiyi Shang}, \bibinfo{person}{Mark~D. Syer},
  {and} \bibinfo{person}{Ahmed~E. Hassan}.} \bibinfo{year}{2018}\natexlab{}.
\newblock \showarticletitle{Examining the Stability of Logging Statements}.
\newblock \bibinfo{journal}{\emph{Empirical Software Engineering}}
  \bibinfo{volume}{23} (\bibinfo{year}{2018}), \bibinfo{pages}{290–333}.
\newblock


\bibitem[\protect\citeauthoryear{Kingma and Ba}{Kingma and Ba}{2015}]%
        {adam}
\bibfield{author}{\bibinfo{person}{Diederik~P. Kingma} {and}
  \bibinfo{person}{Jimmy Ba}.} \bibinfo{year}{2015}\natexlab{}.
\newblock \bibinfo{title}{Adam: A Method for Stochastic Optimization}.
\newblock   (\bibinfo{year}{2015}).
\newblock


\bibitem[\protect\citeauthoryear{Li, Chen, Shang, and Hassan}{Li
  et~al\mbox{.}}{2018a}]%
        {SoftwareLoggingTopicModels}
\bibfield{author}{\bibinfo{person}{Heng Li}, \bibinfo{person}{Tse-Hsun~(Peter)
  Chen}, \bibinfo{person}{Weiyi Shang}, {and} \bibinfo{person}{Ahmed~E.
  Hassan}.} \bibinfo{year}{2018}\natexlab{a}.
\newblock \showarticletitle{Studying Software Logging Using Topic Models}.
\newblock \bibinfo{journal}{\emph{Empirical Softw. Engg.}}
  \bibinfo{volume}{23} (\bibinfo{year}{2018}), \bibinfo{pages}{2655–2694}.
\newblock


\bibitem[\protect\citeauthoryear{Li, Shang, Adams, Sayagh, and Hassan}{Li
  et~al\mbox{.}}{2020b}]%
        {Hassan2020}
\bibfield{author}{\bibinfo{person}{Heng Li}, \bibinfo{person}{Weiyi Shang},
  \bibinfo{person}{Brayan Adams}, \bibinfo{person}{Mohammed Sayagh}, {and}
  \bibinfo{person}{Ahmed~E. Hassan}.} \bibinfo{year}{2020}\natexlab{b}.
\newblock \showarticletitle{A Qualitative Study of the Benefits and Costs of
  Logging from Developers' Perspectives}.
\newblock \bibinfo{journal}{\emph{IEEE Transactions on Software Engineering}}
  \bibinfo{volume}{47} (\bibinfo{year}{2020}), \bibinfo{pages}{2858--2873}.
\newblock
\urldef\tempurl%
\url{https://doi.org/10.1109/TSE.2020.2970422}
\showDOI{\tempurl}


\bibitem[\protect\citeauthoryear{Li, Shang, and Hassan}{Li
  et~al\mbox{.}}{2017}]%
        {Hassan2018}
\bibfield{author}{\bibinfo{person}{Heng Li}, \bibinfo{person}{Weiyi Shang},
  {and} \bibinfo{person}{Ahmed~E. Hassan}.} \bibinfo{year}{2017}\natexlab{}.
\newblock \showarticletitle{Which Log Level Should Developers Choose for a New
  Logging Statement?}
\newblock \bibinfo{journal}{\emph{Empirical Softw. Engg.}}
  \bibinfo{volume}{22}, \bibinfo{number}{4} (\bibinfo{year}{2017}),
  \bibinfo{pages}{1684–1716}.
\newblock


\bibitem[\protect\citeauthoryear{Li, Wang, Lyu, and King}{Li
  et~al\mbox{.}}{2018b}]%
        {codecomp}
\bibfield{author}{\bibinfo{person}{Jian Li}, \bibinfo{person}{Yue Wang},
  \bibinfo{person}{Michael~R. Lyu}, {and} \bibinfo{person}{Irwin King}.}
  \bibinfo{year}{2018}\natexlab{b}.
\newblock \showarticletitle{Code Completion with Neural Attention and Pointer
  Networks}. In \bibinfo{booktitle}{\emph{Proceedings of the 27th International
  Joint Conference on Artificial Intelligence}}. \bibinfo{publisher}{AAAI
  Press}, \bibinfo{pages}{4159–25}.
\newblock


\bibitem[\protect\citeauthoryear{Li}{Li}{2020}]%
        {zhenhao2020}
\bibfield{author}{\bibinfo{person}{Zhenhao Li}.}
  \bibinfo{year}{2020}\natexlab{}.
\newblock \bibinfo{booktitle}{\emph{Towards Providing Automated Supports to
  Developers on Writing Logging Statements}}.
\newblock \bibinfo{publisher}{Association for Computing Machinery},
  \bibinfo{address}{New York, NY, USA}, \bibinfo{pages}{198–201}.
\newblock
\urldef\tempurl%
\url{https://doi.org/10.1145/3377812.3381385}
\showDOI{\tempurl}


\bibitem[\protect\citeauthoryear{Li, Chen, and Shang}{Li
  et~al\mbox{.}}{2020a}]%
        {Li2020}
\bibfield{author}{\bibinfo{person}{Zhenhao Li},
  \bibinfo{person}{Tse-Hsun~(Peter) Chen}, {and} \bibinfo{person}{Weiyi
  Shang}.} \bibinfo{year}{2020}\natexlab{a}.
\newblock \showarticletitle{Where Shall We Log? Studying and Suggesting Logging
  Locations in Code Blocks}. In \bibinfo{booktitle}{\emph{Proceedings of the
  35th IEEE/ACM International Conference on Automated Software Engineering}}.
  \bibinfo{publisher}{Association for Computing Machinery},
  \bibinfo{address}{New York, NY, USA}, \bibinfo{pages}{361–372}.
\newblock


\bibitem[\protect\citeauthoryear{Li, Chen, Yang, and Shang}{Li
  et~al\mbox{.}}{2021a}]%
        {Li2021}
\bibfield{author}{\bibinfo{person}{Zhenhao Li}, \bibinfo{person}{Tse-Hsun~Peter
  Chen}, \bibinfo{person}{Jinqiu Yang}, {and} \bibinfo{person}{Weiyi Shang}.}
  \bibinfo{year}{2021}\natexlab{a}.
\newblock \showarticletitle{Studying Duplicate Logging Statements and Their
  Relationships with Code Clones}.
\newblock \bibinfo{journal}{\emph{IEEE Transactions on Software Engineering}}
  (\bibinfo{year}{2021}).
\newblock
\urldef\tempurl%
\url{https://doi.org/10.1109/TSE.2021.3060918}
\showDOI{\tempurl}


\bibitem[\protect\citeauthoryear{Li, Li, Chen, and Shang}{Li
  et~al\mbox{.}}{2021b}]%
        {DeepLV}
\bibfield{author}{\bibinfo{person}{Zhenhao Li}, \bibinfo{person}{Heng Li},
  \bibinfo{person}{Tse-Hsun Chen}, {and} \bibinfo{person}{Weiyi Shang}.}
  \bibinfo{year}{2021}\natexlab{b}.
\newblock \showarticletitle{DeepLV: Suggesting Log Levels Using Ordinal Based
  Neural Networks}. In \bibinfo{booktitle}{\emph{2021 IEEE/ACM 43rd
  International Conference on Software Engineering (ICSE)}}.
  \bibinfo{publisher}{IEEE Press}, \bibinfo{address}{NJ, USA},
  \bibinfo{pages}{1461--1472}.
\newblock
\urldef\tempurl%
\url{https://doi.org/10.1109/ICSE43902.2021.00131}
\showDOI{\tempurl}


\bibitem[\protect\citeauthoryear{Linnainmaa}{Linnainmaa}{1976}]%
        {Linnainmaa1976}
\bibfield{author}{\bibinfo{person}{Seppo Linnainmaa}.}
  \bibinfo{year}{1976}\natexlab{}.
\newblock \showarticletitle{Taylor expansion of the accumulated rounding
  error}.
\newblock \bibinfo{journal}{\emph{BIT Numerical Mathematics}}
  \bibinfo{volume}{16} (\bibinfo{year}{1976}), \bibinfo{pages}{146--160}.
\newblock


\bibitem[\protect\citeauthoryear{Lundberg}{Lundberg}{2019}]%
        {shapgithub}
\bibfield{author}{\bibinfo{person}{Scott Lundberg}.}
  \bibinfo{year}{2019}\natexlab{}.
\newblock \bibinfo{title}{SHAP Github Implementation}.
\newblock   (\bibinfo{year}{2019}).
\newblock
\urldef\tempurl%
\url{https://github.com/slundberg/shap}
\showURL{%
Retrieved January 2, 2022 from \tempurl}


\bibitem[\protect\citeauthoryear{Lundberg, Erion, Chen, DeGrave, Prutkin, Nair,
  Katz, Himmelfarb, Bansal, and Lee}{Lundberg et~al\mbox{.}}{2020}]%
        {shap}
\bibfield{author}{\bibinfo{person}{Scott~M. Lundberg}, \bibinfo{person}{Gabriel
  Erion}, \bibinfo{person}{Hugh Chen}, \bibinfo{person}{Alex DeGrave},
  \bibinfo{person}{Jordan~M. Prutkin}, \bibinfo{person}{Bala Nair},
  \bibinfo{person}{Ronit Katz}, \bibinfo{person}{Jonathan Himmelfarb},
  \bibinfo{person}{Nisha Bansal}, {and} \bibinfo{person}{Su-In Lee}.}
  \bibinfo{year}{2020}\natexlab{}.
\newblock \showarticletitle{From local explanations to global understanding
  with explainable AI for trees}.
\newblock \bibinfo{journal}{\emph{Nature Machine Intelligence}}
  \bibinfo{volume}{2}, \bibinfo{number}{1} (\bibinfo{year}{2020}),
  \bibinfo{pages}{56--67}.
\newblock


\bibitem[\protect\citeauthoryear{Pecchia, Cinque, Carrozza, and
  Cotroneo}{Pecchia et~al\mbox{.}}{2015}]%
        {Pecchia2015}
\bibfield{author}{\bibinfo{person}{Antonio Pecchia}, \bibinfo{person}{Marcello
  Cinque}, \bibinfo{person}{Gabriella Carrozza}, {and}
  \bibinfo{person}{Domenico Cotroneo}.} \bibinfo{year}{2015}\natexlab{}.
\newblock \showarticletitle{Industry Practices and Event Logging: Assessment of
  a Critical Software Development Process}. In \bibinfo{booktitle}{\emph{2015
  IEEE/ACM 37th IEEE International Conference on Software Engineering}},
  Vol.~\bibinfo{volume}{2}. \bibinfo{publisher}{IEEE Press},
  \bibinfo{address}{New York, USA}, \bibinfo{pages}{169--178}.
\newblock


\bibitem[\protect\citeauthoryear{QOS}{QOS}{2022}]%
        {SLF4J}
\bibfield{author}{\bibinfo{person}{QOS}.} \bibinfo{year}{2022}\natexlab{}.
\newblock \bibinfo{title}{Simple Logging Faced for Java}.
\newblock   (\bibinfo{year}{2022}).
\newblock
\urldef\tempurl%
\url{https://www.slf4j.org/}
\showURL{%
Retrieved January 2, 2022 from \tempurl}


\bibitem[\protect\citeauthoryear{Shang, Jiang, Adams, Hassan, Godfrey, Nasser,
  and Flora}{Shang et~al\mbox{.}}{2014}]%
        {Sheng2014}
\bibfield{author}{\bibinfo{person}{Weiyi Shang}, \bibinfo{person}{Zhen~Ming
  Jiang}, \bibinfo{person}{Bram Adams}, \bibinfo{person}{Ahmed~E. Hassan},
  \bibinfo{person}{Michael~W. Godfrey}, \bibinfo{person}{Mohamed Nasser}, {and}
  \bibinfo{person}{Parminder Flora}.} \bibinfo{year}{2014}\natexlab{}.
\newblock \showarticletitle{An exploratory study of the evolution of
  communicated information about the execution of large software systems}.
\newblock \bibinfo{journal}{\emph{Journal of Software: Evolution and Process}}
  \bibinfo{volume}{26} (\bibinfo{year}{2014}), \bibinfo{pages}{3--26}.
\newblock


\bibitem[\protect\citeauthoryear{Shilin, Jieming, Pinjia, and Michael}{Shilin
  et~al\mbox{.}}{2016}]%
        {PinjiaHe2016}
\bibfield{author}{\bibinfo{person}{He Shilin}, \bibinfo{person}{Zhu Jieming},
  \bibinfo{person}{He Pinjia}, {and} \bibinfo{person}{Lyu Michael, R.}}
  \bibinfo{year}{2016}\natexlab{}.
\newblock \showarticletitle{Experience Report: System Log Analysis for Anomaly
  Detection}. In \bibinfo{booktitle}{\emph{2016 IEEE 27th International
  Symposium on Software Reliability Engineering (ISSRE)}}.
  \bibinfo{publisher}{IEEE Press}, \bibinfo{address}{River Side, USA},
  \bibinfo{pages}{207--218}.
\newblock


\bibitem[\protect\citeauthoryear{Sim and Wright}{Sim and Wright}{2005}]%
        {highKappa}
\bibfield{author}{\bibinfo{person}{Julius Sim} {and} \bibinfo{person}{Chris~C
  Wright}.} \bibinfo{year}{2005}\natexlab{}.
\newblock \showarticletitle{{The Kappa Statistic in Reliability Studies: Use,
  Interpretation, and Sample Size Requirements}}.
\newblock \bibinfo{journal}{\emph{Physical Therapy}}  \bibinfo{volume}{85}
  (\bibinfo{year}{2005}), \bibinfo{pages}{257--268}.
\newblock


\bibitem[\protect\citeauthoryear{Tony}{Tony}{1995}]%
        {distributedRepresentations}
\bibfield{author}{\bibinfo{person}{A.~{Plate} Tony}.}
  \bibinfo{year}{1995}\natexlab{}.
\newblock \showarticletitle{Holographic reduced representations}.
\newblock \bibinfo{journal}{\emph{IEEE Transactions on Neural Networks}}
  \bibinfo{volume}{6}, \bibinfo{number}{3} (\bibinfo{year}{1995}),
  \bibinfo{pages}{623--641}.
\newblock
\urldef\tempurl%
\url{https://doi.org/10.1109/72.377968}
\showDOI{\tempurl}


\bibitem[\protect\citeauthoryear{van~der Aalst and al}{van~der Aalst and
  al}{2012}]%
        {manifesto}
\bibfield{author}{\bibinfo{person}{van~der Aalst} {and} \bibinfo{person}{et
  al}.} \bibinfo{year}{2012}\natexlab{}.
\newblock \showarticletitle{Process Mining Manifesto}. In
  \bibinfo{booktitle}{\emph{Business Process Management Workshops}}.
  \bibinfo{publisher}{Springer Berlin Heidelberg}, \bibinfo{address}{Berlin,
  Heidelberg}, \bibinfo{pages}{169--194}.
\newblock


\bibitem[\protect\citeauthoryear{Vaswani, Shazeer, Parmar, Uszkoreit, Jones,
  Gomez, Kaiser, and Polosukhin}{Vaswani et~al\mbox{.}}{2017}]%
        {Transformer}
\bibfield{author}{\bibinfo{person}{Ashish Vaswani}, \bibinfo{person}{Noam
  Shazeer}, \bibinfo{person}{Niki Parmar}, \bibinfo{person}{Jakob Uszkoreit},
  \bibinfo{person}{Llion Jones}, \bibinfo{person}{Aidan~N. Gomez},
  \bibinfo{person}{undefinedukasz Kaiser}, {and} \bibinfo{person}{Illia
  Polosukhin}.} \bibinfo{year}{2017}\natexlab{}.
\newblock \showarticletitle{Attention is All You Need}. In
  \bibinfo{booktitle}{\emph{Proceedings of the 31st International Conference on
  Neural Information Processing Systems}}. \bibinfo{publisher}{Curran
  Associates Inc.}, \bibinfo{address}{Red Hook, NY, USA},
  \bibinfo{pages}{6000–6010}.
\newblock


\bibitem[\protect\citeauthoryear{Watters and Boslaugh}{Watters and
  Boslaugh}{2008}]%
        {StatNut}
\bibfield{author}{\bibinfo{person}{Andrew Watters} {and} \bibinfo{person}{Sarah
  Boslaugh}.} \bibinfo{year}{2008}\natexlab{}.
\newblock \bibinfo{booktitle}{\emph{Statistics in a Nutshell: A Desktop Quick
  Reference}}.
\newblock \bibinfo{publisher}{O’Reilly Media}, \bibinfo{address}{USA}.
\newblock


\bibitem[\protect\citeauthoryear{Yuan, Park, Huang, Liu, Lee, Tang, Zhou, and
  Savage}{Yuan et~al\mbox{.}}{2012b}]%
        {ErrLog}
\bibfield{author}{\bibinfo{person}{Ding Yuan}, \bibinfo{person}{Soyeon Park},
  \bibinfo{person}{Peng Huang}, \bibinfo{person}{Yang Liu},
  \bibinfo{person}{Michael~M. Lee}, \bibinfo{person}{Xiaoming Tang},
  \bibinfo{person}{Yuanyuan Zhou}, {and} \bibinfo{person}{Stefan Savage}.}
  \bibinfo{year}{2012}\natexlab{b}.
\newblock \showarticletitle{Be Conservative: Enhancing Failure Diagnosis with
  Proactive Logging}. In \bibinfo{booktitle}{\emph{10th {USENIX} Symposium on
  Operating Systems Design and Implementation ({OSDI} 12)}}.
  \bibinfo{publisher}{{USENIX} Association}, \bibinfo{address}{Hollywood, CA},
  \bibinfo{pages}{293--306}.
\newblock


\bibitem[\protect\citeauthoryear{Yuan, Park, and Zhou}{Yuan
  et~al\mbox{.}}{2012a}]%
        {Yuan2012}
\bibfield{author}{\bibinfo{person}{Ding Yuan}, \bibinfo{person}{Soyeon Park},
  {and} \bibinfo{person}{Yuanyuan Zhou}.} \bibinfo{year}{2012}\natexlab{a}.
\newblock \showarticletitle{Characterizing Logging Practices in Open-Source
  Software}. In \bibinfo{booktitle}{\emph{Proceedings of the 34th International
  Conference on Software Engineering}}. \bibinfo{publisher}{IEEE Press},
  \bibinfo{address}{River Street, USA}, \bibinfo{pages}{102–112}.
\newblock


\bibitem[\protect\citeauthoryear{Yuan, Zheng, Park, Zhou, and Savage}{Yuan
  et~al\mbox{.}}{2012c}]%
        {LogEnchancer}
\bibfield{author}{\bibinfo{person}{Ding Yuan}, \bibinfo{person}{Jing Zheng},
  \bibinfo{person}{Soyeon Park}, \bibinfo{person}{Yuanyuan Zhou}, {and}
  \bibinfo{person}{Stefan Savage}.} \bibinfo{year}{2012}\natexlab{c}.
\newblock \showarticletitle{Improving Software Diagnosability via Log
  Enhancement}.
\newblock \bibinfo{journal}{\emph{ACM Trans. Comput. Syst.}}
  \bibinfo{volume}{30}, \bibinfo{number}{1} (\bibinfo{year}{2012}),
  \bibinfo{pages}{3–14}.
\newblock


\bibitem[\protect\citeauthoryear{Zeng, Chen, Shang, and Chen}{Zeng
  et~al\mbox{.}}{2019}]%
        {Zeng2019}
\bibfield{author}{\bibinfo{person}{Yi Zeng}, \bibinfo{person}{Jinfu Chen},
  \bibinfo{person}{Weiyi Shang}, {and} \bibinfo{person}{Tse-Hsun~(Peter)
  Chen}.} \bibinfo{year}{2019}\natexlab{}.
\newblock \showarticletitle{Studying the characteristics of logging practices
  in mobile apps: a case study on F-Droid}.
\newblock \bibinfo{journal}{\emph{Empirical Software Engineering}}
  \bibinfo{volume}{24} (\bibinfo{year}{2019}), \bibinfo{pages}{3394--3434}.
\newblock


\bibitem[\protect\citeauthoryear{Zhao, Rodrigues, Luo, Stumm, Yuan, and
  Zhou}{Zhao et~al\mbox{.}}{2017}]%
        {20Log}
\bibfield{author}{\bibinfo{person}{Xu Zhao}, \bibinfo{person}{Kirk Rodrigues},
  \bibinfo{person}{Yu Luo}, \bibinfo{person}{Michael Stumm},
  \bibinfo{person}{Ding Yuan}, {and} \bibinfo{person}{Yuanyuan Zhou}.}
  \bibinfo{year}{2017}\natexlab{}.
\newblock \showarticletitle{Log20: Fully Automated Optimal Placement of Log
  Printing Statements under Specified Overhead Threshold}. In
  \bibinfo{booktitle}{\emph{Proceedings of the 26th Symposium on Operating
  Systems Principles}}. \bibinfo{publisher}{Association for Computing
  Machinery}, \bibinfo{address}{New York, NY, USA}, \bibinfo{pages}{565–581}.
\newblock


\bibitem[\protect\citeauthoryear{Zhi, Yin, Deng, Ye, Fu, and Xie}{Zhi
  et~al\mbox{.}}{2019}]%
        {taoxie}
\bibfield{author}{\bibinfo{person}{Chen Zhi}, \bibinfo{person}{Jianwei Yin},
  \bibinfo{person}{Shuiguang Deng}, \bibinfo{person}{Maoxin Ye},
  \bibinfo{person}{Min Fu}, {and} \bibinfo{person}{Tao Xie}.}
  \bibinfo{year}{2019}\natexlab{}.
\newblock \showarticletitle{An Exploratory Study of Logging Configuration
  Practice in Java}. In \bibinfo{booktitle}{\emph{2019 IEEE International
  Conference on Software Maintenance and Evolution (ICSME)}}.
  \bibinfo{publisher}{IEEE Press}, \bibinfo{address}{New York, USA},
  \bibinfo{pages}{459--469}.
\newblock
\urldef\tempurl%
\url{https://doi.org/10.1109/ICSME.2019.00079}
\showDOI{\tempurl}


\bibitem[\protect\citeauthoryear{Zhu, He, Fu, Zhang, Lyu, and Zhang}{Zhu
  et~al\mbox{.}}{2015}]%
        {LogAdvisor}
\bibfield{author}{\bibinfo{person}{Jieming Zhu}, \bibinfo{person}{Pinjia He},
  \bibinfo{person}{Qiang Fu}, \bibinfo{person}{Hongyu Zhang},
  \bibinfo{person}{Michael~R. Lyu}, {and} \bibinfo{person}{Dongmei Zhang}.}
  \bibinfo{year}{2015}\natexlab{}.
\newblock \showarticletitle{Learning to Log: Helping Developers Make Informed
  Logging Decisions}. In \bibinfo{booktitle}{\emph{Proceedings of the 37th
  International Conference on Software Engineering - Volume 1}}.
  \bibinfo{publisher}{IEEE Press}, \bibinfo{address}{River Street, USA},
  \bibinfo{pages}{415–425}.
\newblock


\end{thebibliography}

\appendix
\section{Sample analysis: When QuLog outperforms the baselines?}~\label{examp}
One set of examples where we find that QuLog's log level prediction outperforms other methods is in the logs like "created with buffersize=* and maxpoolsize=*" (with original log level INFO, QuLog prediction is "info", "DeepLV" and "BERT\_SVM" are predicting "warning") and "stopping the wall procedure store, isabort= | (self aborting)" (with original log level INFO, QuLog prediction is "info", "DeepLV" and "BERT\_SVM" are predicting "error"). These two messages contain words like "maxpoolsize", "buffersize" or "isabort" which are not standard in general language. Therefore, the log representation by general language models is inferior in comparison to QuLog, which directly learns log representations from the static text. Since QuLog, is directly trained on log-specific words it can learn better representations and relate the words to the correct target. 

When considering the linguistic comparison, as seen by the results, QuLog and BERT\_SVM are performing similarly. We find one consistent example where QuLog performs better than SVM, i.e., when HBase is used for testing QuLog correctly predicts the linguistic group "VERB VERB" as insufficient. We hypothesise that QuLog can extrapolate this information having the groups "VERB" and "VERB NOUN VERB" as insufficient during training. The two baselines, SVM and RF are performing slightly differently on the same representation, we consider the observed difference to reside in the type of decision boundary they fit. SVMs are characterized to perform better in very high dimensional spaces (because of the properties of such spaces). RF is characterized by diverse volume pockets in different regions. Since the samples in the high dimensions are very far apart from one another the pocket regions created by RF lead to more miss-classifications.

\section{Comparing QuLog linguistic module against rule-based approach}~\label{rule-basedapproach}
The process of identification of sufficient linguistic structure was conducted as follows. Two human annotators examined the linguistic structure and the raw static texts organized by linguistic groups. A linguistic group is a set of static text instructions'. A linguistic group is identified by the sequence of POS tags. Each human annotator examined the individual static text within each group (within each of the randomly sampled 361 groups). Since each log message should describe an event verbosely and convey sufficient information on one side, and following the maxim of quality and quantity for short texts from general language properties, on the other side, the static text should have minimal linguistic structure for sufficient expression of the information. By examining the raw static text within each group the annotators relate their understanding of the sufficient information a log message has. They assign labels 0 for "linguistically sufficient" or 1 for the "linguistically insufficient" group. Then the labels provided by the two human annotators are compared, and the linguistic groups with overlapping labels are retained. Each of the static texts is then labeled with the label of the linguistic group associated with it. However, the model is trained using the linguistic group representation of the static text obtained by the pos tag. The labels can be used as rules what is a linguistically good and bad static text.

The generated rules can be used to detect logs with sufficient and insufficient structure (that is how they are generated). Whenever we have a new system, we calculate their linguistic representations and match them against the rules. Any match is considered as a log with insufficient quality. For example, if we find a static text with "NOUN NOUN" from the new system, we can say that the static text is of "insufficient" quality. However, we found examples, where QuLog maybe interpolates between two nearby rules, and correctly identify a rule from a new project that was not part of the training data. For example, when HBase is in the test set, (the rules in the training set are extracted from other systems), the rules do not cover the linguistically insufficient group "VERB VERB". In contrast, QuLog correctly predicts this group as "insufficient" because of its similarities with the linguistically insufficient groups of "VERB" and 'VERB NOUN VERB' (that are slightly different from "VERB VERB"). Since QuLog has strong performance, and can interpolate between rules we opt for the given design. Nevertheless, there is still much space for improving the input training dataset that will lead to new insights.

\section{Linguistic Quality Assessment Additional Evidence}~\label{additionalEvidence}
The idea for "sufficient linguistic quality" assesment is built around Jira issues like ZOOKEEPER-2126, ZOOKEEPER-3659 and Zookeeper-259. They show that the minimal information in the static text hurts comprehensibility. By relating the information about minimality (e.g., minimal number of tokens, linguistic token categories) with the (in)sufficient linguistic structure of the static text we can evaluate the type of linguistic categories (e.g., verb, nouns, adjectives, and similar) participating in the verbose and comprehensive description of the events. By enriching the event description with additional linguistic properties, the log messages are easier for reading, comprehension, and contain sufficient verbose information. 

In the following, we further describe two Jira issues related to linguistic structural problems. In  \href{https://issues.apache.org/jira/browse/ZOOKEEPER-3659}{ZOOKEEPER-3659} it is reported that WatchManagerFactory log is not sufficiently readable. The fix of this issue is to change the prior static text \textit{Using org.apache.zookeeper.server.watch.WatchManager as watch manager}, into \textit{dataWatches is using org.apache.zookeeper.server.watch.WatchManager as watch manager.}. From the linguistic perspective this means that the linguistic structure ['VERB', 'PUNCT', 'ADP', 'NOUN', 'NOUN'] is transformed into [\textbf{'NOUN', 'AUX', } 'VERB', 'PUNCT', 'ADP', 'NOUN', 'NOUN']. The additional linguistic concepts improve the comprehensibility of the event.

Similarly in the Jira issue \href{https://issues.apache.org/jira/browse/ZOOKEEPER-259}{Zookeeper-259} despite the correction of the log levels, the text in the log instructions is changed as well. One example is the change on lines 524, through 526. Specifically, the prior static text "Got ping sessionid:0x" was replaced with "Got ping response for sessionid:0x". Linguistically speaking this means that the text is changed from ['AUX', 'VERB', 'NOUN'] into ['VERB', 'NOUN', 'NOUN', 'ADP', 'NOUN'] which improves the comprehensibility and makes the log line easer for the operators to understand.

\end{document}